\documentclass{ws-mpla}
\usepackage{ulem}
\usepackage[super]{cite}
\usepackage{xcolor}
\usepackage[verbose,hypertexnames=false]{hyperref}
\definecolor{blue}{rgb}{0.0, 0.0, 1.0}
\definecolor{red}{rgb}{1.0, 0.0, 0.0}
\definecolor{royalblue}{rgb}{0.0, 0.14, 0.4}
\hypersetup{colorlinks=true,linkcolor=red, citecolor=blue, urlcolor=blue}
\def\orcid#1{\kern .08em\href{https://orcid.org/#1}{\includegraphics[keepaspectratio,width=0.7em]{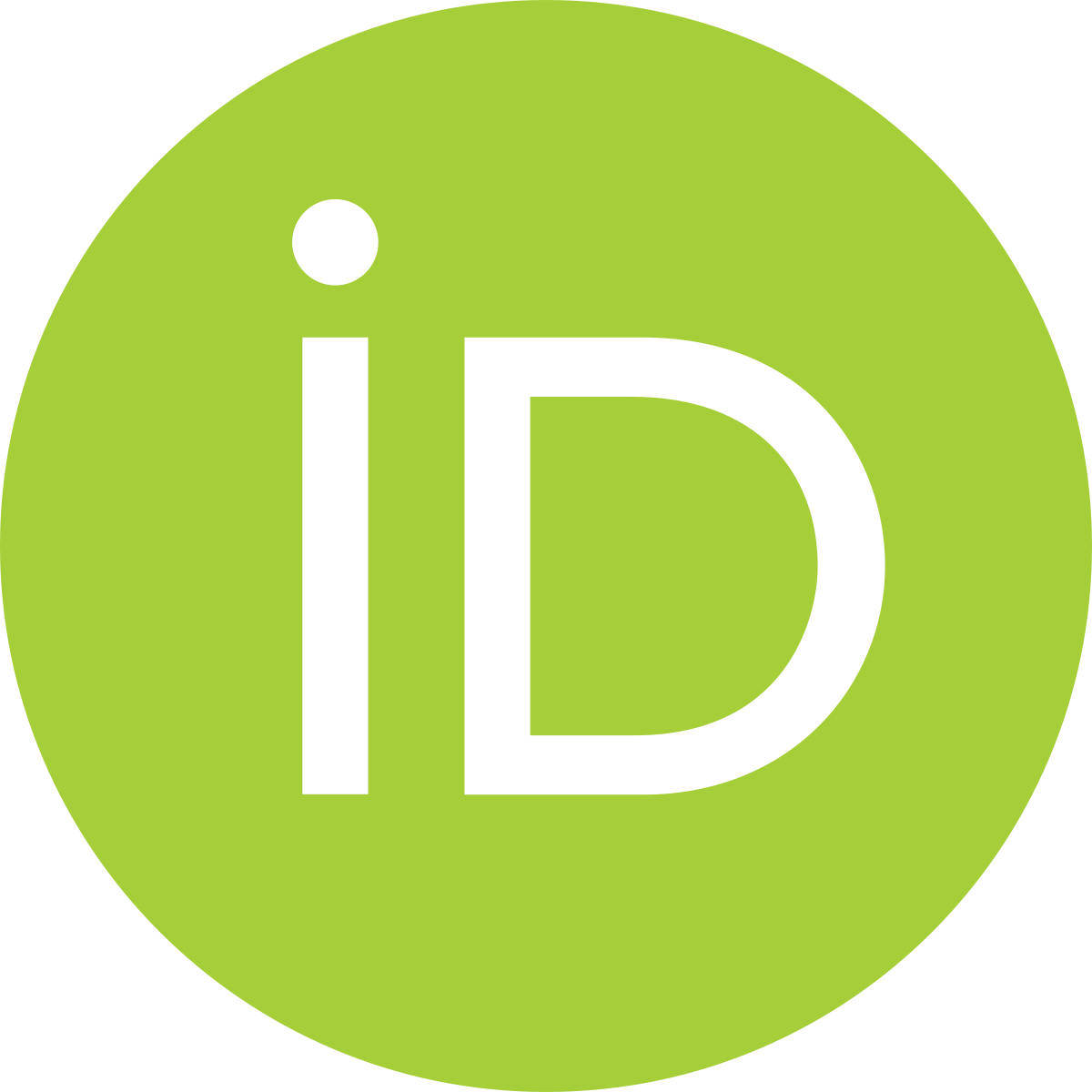}}}

\begin{document}

\markboth{Fernando Chandra \textit{et al.,}}{Pion generalized parton distributions at zero skewness}

\catchline{}{}{}{}{}
\title{Pion generalized parton distributions at zero skewness}

\author{Fernando Chandra\orcid{0009-0003-8583-4054}}
\address{Departemen Fisika, FMIPA, Universitas Indonesia\\
Depok, 16424, Indonesia\\
E-Mail Address: fernando.chandra@ui.ac.id}
\author{Parada T.~P.~Hutauruk\orcid{0000-0002-4225-7109}}
\address{Department of Physics, Pukyong National University (PKNU)\\
Busan, 48513, Korea \\
Departemen Fisika, FMIPA, Universitas Indonesia\\
Depok, 16424, Indonesia\\
E-Mail Address: phutauruk@gmail.com}
\author{Terry Mart\orcid{0000-0003-4628-2245}}
\address{Departemen Fisika, FMIPA, Universitas Indonesia\\
Depok, 16424, Indonesia\\
E-Mail Address: terry.mart@sci.ui.ac.id}
\maketitle

\begin{history}
\received{(Day Month Year)}
\revised{(Day Month Year)}
\accepted{(Day Month Year)}
\published{(Day Month Year)}
\end{history}

\begin{abstract}
In this paper, we systematically study the generalized parton distributions (GPDs) of the pion Goldstone Boson
at zero skewness ($\xi =0$) in the framework of the covariant Nambu--Jona-Lasinio model with the help of the proper time regularization scheme to cure the divergence simultaneously to simulate the confinement. To this end, we evaluate the generalized form factors and parton distribution functions derived, respectively, from the first Mellin moments and the forward limit pion GPDs, in comparison to existing experimental data, recent lattice QCD simulations, and JAM global QCD analyses. We find that the pion parton distribution functions derived from the pion GPDs have excellent agreement with the experimental data and JAM analysis at renormalization scale $\mu^2 =$ 4 and 27 GeV$^2$. In addition, our results for the pion generalized form factors involving the scalar, vector, and tensor dressed form factors are consistent with recent lattice data and existing data. We then compute the charge radii for those dressed generalized form factors, and we obtain $r_{S}^{\pi} =$ 0.56 fm, $ r_V^{\pi} =$ 0.63 fm, and $r_{T}^{\pi} =$ 0.83 fm for the pion scalar, vector, and tensor form factors, respectively.
\end{abstract}

\keywords{Nambu--Jona-Lasinio model; generalized parton distributions; proper-time regularization scheme, pseudo-Goldstone boson; 
generalized form factors.}

\section{Introduction}	
The internal structure of hadrons, as a playground of quantum chromodynamics (QCD)~\cite{Gross:2022hyw}, provides crucial information about fundamental properties of QCD, such as spontaneous chiral symmetry breaking~\cite{Lee:1972fj}, asymptotic freedom~\cite{Politzer:1974fr,Gross:1973ju,Gross:1973zrg}, and confinement. Among hadrons, the pion occupies a special role as a pseudo-Goldstone boson associated with the spontaneous breaking of chiral symmetry in QCD, as well as a bound state of the strong interaction. These features make the pion an interesting and challenging object for further study. However, the internal structure of the pion cannot yet be directly computed from the first principles of QCD, despite the fact that QCD is believed to be the underlying theory of the strong interaction. 
To overcome this limitation, theoretical frameworks that mimic essential features of QCD, commonly referred to as QCD-inspired models, have been developed. In the literature, a variety of QCD-inspired models have been employed to investigate the internal structure of the pion and to gain insight into the properties of QCD~\cite{Adhikari:2021jrh,Frederico:2009fk,Kaur:2025gyr,Diakonov:1997sj,Praszalowicz:2001wy,Hutauruk:2025wkn,Son:2024uet,Hutauruk:2023ccw,Liu:2024jno,Kwee:2007dd,Abidin:2019xwu,Hutauruk:2016sug,Chang:2014gga}. Within these QCD-inspired models, it is possible to directly compute multidimensional distribution functions, including generalized parton distributions (GPDs), parton distribution functions (PDFs), electromagnetic elastic form factors (EMFFs), and generalized transverse momentum-dependent (GTMD) parton distributions, through matrix element or quark correlator functions. These observables provide deeper insight into the internal structure of hadrons, specifically their spatial, transverse, and longitudinal momentum fractions of the valence quark inside the pion. However, in the present study, we focus on pion GPDs at zero skewness, which provides information about the quark longitudinal momentum fraction, skewness parameter, and transverse momentum with respect to the pion parent momentum.

The GPDs are powerful and essential tools for understanding the internal structure of the pion in three-dimensional (3D) tomography. They unify the information contained in PDFs and EMFFs. In this sense, pion GPDs provide access to one-dimensional PDFs and EMFFs, in addition to a three-dimensional tomographic description of the pion. Unfortunately, experimental data for the pion GPDs are not yet available. However, limited data exist for the pion PDFs, and several measurements of pion EMFFs have been reported. Therefore, in this work, we extract pion PDFs and EMFFs from the pion GPDs and compare them with existing experimental data~\cite{E615:1989bda} and with the JAM QCD analysis~\cite{Barry:2021osv}. It is worth noting that the pion PDFs must be evolved using QCD evolution equations before meaningful comparison with experimental data can be made. In addition, the results of this study will be relevant for comparison with data from future experiments, such as the Electron-Ion Collider (EIC) at BNL~\cite{Arrington:2021biu}, the Electron-ion collider in China (EicC)~\cite{Anderle:2021wcy}, J-PARC~\cite{Sawada:2016mao}, the upgraded JLAB 22 GeV~\cite{Accardi:2023chb}, and the COMPASS/AMBER++ at CERN~\cite{Adams:2018pwt}.

To achieve these purposes, we systematically investigate the pion GPDs at zero skewness within the framework of the covariant Nambu-Jona-Lasinio (NJL) model, incorporating the proper time regularization scheme that cures the loop divergence and simultaneously simulates quark confinement. In applying the proper-time regularization scheme, we introduce both UV and IR cutoffs in the calculation. The UV cutoff is required to render the theory finite, while the IR cutoff is introduced to remove the unphysical quark-production threshold that would otherwise allow the meson to decay into free quarks. In this scheme, the IR cutoff effectively mimics the QCD confinement scale. We first compute and predict the vector and tensor pion GPDs. We then extract the pion PDFs and EMFFs from the pion GPDs at zero skewness and compare them with existing experimental data~\cite{E615:1989bda} and the JAM global analysis~\cite{Barry:2021osv}. As noted earlier, to compare the pion PDFs calculated at a low-energy scale with experimental data at higher energy scales, the PDFs must be evolved using the DGLAP QCD evolution equations~\cite{Miyama:1995bd}. Accordingly, we evolve the pion PDFs from the initial model scale $\mu_0^2 =$ 0.18 GeV$^2$ to $\mu^2 =$ 27 and 4 GeV$^2$, which are the scales used in the experiments and lattice QCD, respectively. Here, we also compute the scalar, vector, and tensor generalized form factors of the pion. Specifically, the scalar form factor is derived from the twist-3 scalar pion GPD, while the vector and tensor form factors are obtained from the $n=0$ Mellin moments of the vector and tensor GPDs, respectively. Then, we compare these results with recent lattice QCD calculations~\cite{Alexandrou:2021ztx}. In addition, we compute and present results for the charge radii associated with the scalar, vector, and tensor pion form factors.

This paper is organized as follows. We present a brief introduction of the covariant NJL model with the proper time regularization scheme, including the calculations of the gap mass, the pion-quark coupling, and pion mass, in Sec.~\ref{sec:NJL}. Section~\ref{sec:gpd} presents the generic formulation and properties of GPDs and the pion GPDs in the covariant NJL model, including the relation of the pion GPDs and the pion PDFs and EMFFs. Additionally, we show how to compute the scalar, vector, and tensor pion form factors from the twist-3 scalar pion GPDs and the first Mellin moments of the zero skewness pion GPDs, respectively. We present and discuss the numerical results for the pion GPDs, PDFs, and scalar, vector, and tensor pion form factors, together with their corresponding charge radii in Sec.~\ref{sec:num}. Summary and conclusion of this work are given in Sec.~\ref{sec:sum}.

\section{NJL model framework} \label{sec:NJL}
In this section, we briefly introduce the SU(2) flavor NJL model. The NJL model has been successfully applied to a wide range of physical phenomena, including pion and kaon PDFs~\cite{Hutauruk:2016sug}, kaon EMFFs~\cite{Ninomiya:2014kja}, charge symmetry violation in PDFs and EMFFs~\cite{Hutauruk:2018zfk}, fragmentation function~\cite{Bentz:2016rav}, rho meson TMDs and EMFFs~\cite{Ninomiya:2017ggn,Carrillo-Serrano:2015uca,Hutauruk:2025bjd}, as well as nuclear matter and neutron-star properties~\cite{Bentz:2001vc,Noro:2023vkx}. 
The Lagrangian of the SU(2) flavor NJL model is given by
\begin{eqnarray}
    \mathcal{L}_{\mathrm{NJL}} &=& \bar{\psi}_q \big( i\partial\!\!\!/ - \hat{m}_q) \psi_q +
\frac{G_\pi}{2} \Big[ \big( \bar{\psi}_q \psi_q \big)^2 - \big( \bar{\psi}_q \vec{\tau} \gamma_5 \psi_q
\big)^2 \Big] - \frac{G_\omega}{2}  \big( \bar{\psi}_q \gamma^\mu \psi_q \big)^2 \nonumber \\
&-& \frac{G_\rho}{2}
 \Big[ \big( \bar{\psi}_q \gamma^\mu \vec{\tau} \psi_q \big)^2 + \big( \bar{\psi}_q
 \gamma^\mu \gamma_5 \vec{\tau}\psi_q \big)^2 \Big],
\end{eqnarray}
where $G_\pi$, $G_\omega$, and $G_\rho$ denote the scalar, $\omega$ and $\rho$ mesons coupling constants, respectively. The quark field with flavor $q =(u,d)$ is denoted by $\psi_q$, and $\hat{m}_q$ is the (bare) current quark masses, where $q =(u,d)$. In this work, we assume $m_u =m_d$, corresponding to SU(2) isospin symmetry. The gap equation in the NJL model is determined from the quark self-energy interaction, which gives
\begin{eqnarray}
    \label{eq:gapnjl}
    M_q &=& m_q - 4 G_\pi \big< \bar{\psi}_q \psi_q \big> = m_q + 4 i G_\pi \int \frac{d^4k}{(2\pi)^4} \mathrm{Tr} \big[ S_q (k) \big], 
\end{eqnarray}
where $S_q (k) = {\big(k\!\!\!/ + M_q \big)}/{\big(k^2 -M_q^2 + i \epsilon \big)}$ is the quark propagator in momentum space and $\big< \bar{\psi}_q \psi_q \big>$ stands for the quark condensate, which is the order parameter of the chiral symmetry breaking. After applying the Wick rotation and introducing the proper time regularization scheme, one has
\begin{eqnarray}
    \label{eq:gapptr}
    M_q &=& m_q  + \frac{3G_\pi M_q}{\pi^2} \int_{\tau_{\mathrm{UV}}^2}^{\tau_{\mathrm{IR}}^2} \frac{d\tau}{\tau^2} \exp \big[ -\tau \big( M_q^2 \big)\big],
\end{eqnarray}
where $\tau_{\mathrm{UV}}^2 = 1/\Lambda_{\mathrm{UV}}^2$ and $\tau_{\mathrm{IR}}^2 = 1/\Lambda_{\mathrm{IR}}^2$ are the lower (ultraviolet) and upper (infrared) limits of the quark momentum integrations, respectively. In this work, we set $\Lambda_{\mathrm{IR}} =$ 240 MeV, which is close to $\Lambda_{\mathrm{QCD}} \simeq $ (200-350) MeV. The latter range is typically obtained at the one-loop level of perturbation theory, where the running coupling is given by $\alpha_s(Q) \simeq \big[\ln\left(Q^2/\Lambda_{\mathrm{QCD}}^2\right)\big]^{-1}$. 

Using the random phase approximation (RPA), which is equivalent to the ladder approximation, the pion $T$-matrices can be determined from the quark-antiquark interaction (scattering) in the pion channel. The sum of the infinite bubble diagrams in the RPA can be expressed as
\begin{eqnarray}
    \label{eq:rpa}
    T_{\pi} &=& \gamma_5 \tau_j \frac{-2i G_\pi}{\big[ 1+ 2 G_\pi \Pi_{\pi} (k^2) \big]} \gamma_5 \tau_j,
\end{eqnarray}
where the bubble diagram or the polarization of the quark propagator for the pion is defined as
\begin{eqnarray}
    \label{eqn:bubpion}
    \Pi_\pi (k^2) \delta_{ab} &=& i \int \frac{d^4p}{(2\pi)^4} \mathrm{Tr} \big[ \gamma_5 \tau_a S_q (p+k) \gamma_5 \tau_b S_q (p) \big].
\end{eqnarray}
The pion mass is determined from the pole of the corresponding $T$-matrix. The pole condition is given by
\begin{eqnarray}
    \label{eq:pionmass}
    1 + 2 G_\pi \Pi_{\pi} (k^2 = m_\pi^2) &=& 0,
\end{eqnarray}
where in the vicinity of the $T$-matrix pole associated with the pion bound state, the $T$-matrix can be written as
\begin{eqnarray}
\label{eq:Tmatrixpion}
    T_{\pi} \sim \gamma_5 \tau_j \frac{ig_{\pi qq}^2}{k^2 -m_\pi^2 + i \epsilon} \gamma_5  \tau_j,
\end{eqnarray}
where $g_{\pi qq}$ is the pion-quark coupling constant. By expanding the sum of the bubble diagrams in Eq.~(\ref{eq:rpa}) around the pole at $k^2 = m_\pi^2$, the pion-quark coupling constant can be derived as follows,
\begin{eqnarray}
    \Pi_{\pi} (k^2) &=& \Pi_{\pi} (k^2 = m_\pi^2) + \frac{\partial \Pi_\pi (k^2)}{\partial k^2} \Bigg|_{k^2 = m_\pi^2} \big( k^2 - m_\pi^2 \big) + \cdot \cdot \cdot,
\end{eqnarray}
which leads to
\begin{eqnarray}
    g_{\pi qq}^{-2} &=& - \Bigg( \frac{\partial \Pi_\pi (k^2)}{\partial k^2}\Bigg) \Bigg|_{k^2 = m_\pi^2}.
\end{eqnarray}
The resulting pion mass and the pion-quark coupling constant are then used as inputs for the computation of pion GPDs.

\section{Generalized parton distributions} \label{sec:gpd}
In this section, we present the calculation of the vector, tensor, and scalar pion GPDs within the covariant NJL model, starting from their definitions and generic expressions. In general, GPDs are formulated through nonlocal quark-quark matrix elements in non-diagonal momentum space ($\xi \neq 0$), with the quark fields separated along the light-like direction, and can be defined as follows:
\begin{subequations}
\label{eq:gpd1}
\begin{equation}
        \mathcal{H}^q (x, \xi,t) = \int \frac{dz^{-}}{4\pi } e^{\big[ ixP^+ z^{-}\big]} \big< \pi^+(p') \mid \bar{\psi}_q (-\frac{z^{-}}{2}) \gamma^+ \psi_q (\frac{z^{-}}{2}) \mid \pi^+ (p) \big>_{z^+ =0, \mathbf{z} =0},
\end{equation}
\begin{equation}
    \mathcal{P} E^q (x,\xi,t) = \int \frac{dz^{-}}{4\pi} e^{\big[ ixP^+ z^-\big]} \big< \pi^+(p') \mid \bar{\psi}_q (-\frac{z^{-}}{2}) i \sigma^{+j} \psi_q (\frac{z^{-}}{2}) \mid \pi^+ (p) \big>_{z^+ =0, \mathbf{z} = 0},
\end{equation}
\begin{equation}
    \frac{M_u}{P^+} \mathcal{H}^q_{S} (x,\xi,t) = \int \frac{dz^{-}}{4\pi} e^{\big[ixP^+ z^-\big]} \big< \pi^+ (p') \mid \bar{\psi}_q (- \frac{z^-}{2}) 1 \psi_q (\frac{z^-}{2}) \mid \pi^+ (p) \big>_{z^+ =0, \mathbf{z} =0},
\end{equation}
\end{subequations}
where $\mathcal{P} = {(P^+ q^j - P^j q^+)}/{(m_\pi P^+)}$, $\psi_q$ is the quark field with flavor $q =(u,d)$, $x$ is the Bjorken $x$ or the quark longitudinal momentum, $t= Q^2 = -q^2 = \Delta^2 = (p' -p)^2 $ is the momentum transfer or photon virtual momentum, and $\xi = {(p^+ - p'^{+})}/{(p^+ + p'^+)} = - {\Delta^+}/{2P^+}$ stands for the skewness parameter with the $p$ and $p'$ are the initial and final pion momentum, respectively. The average momentum of the pion is defined by $P = {(p+p')}/{2}$.

The pion has leading twist vector (no spin flip) $\mathcal{H}^q (x, \xi,t)$ and tensor (spin flip) $E^q (x, \xi, t) $ quark GPDs. The vector and tensor quark GPDs can be defined in the isoscalar and isovector isospin projections with the isospin matrix equal unity for the isoscalar and $\tau_3$ for the isovector. The relation between the quark GPDs and the isoscalar and isovector GPDs can be written as
\begin{eqnarray}
    \mathcal{H}^q (x,\xi,t) &=& \frac{1}{2} \big[ \mathcal{H}^{I=0} (x,\xi,t) + \mathcal{H}^{I=1} (x,\xi,t) \big], \\
    \mathcal{H}^{\bar{q}} (x,\xi,t) &=& \frac{1}{2} \big[ \mathcal{H}^{I=0} (x,\xi,t) - \mathcal{H}^{I=1} (x,\xi.t) \big],
\end{eqnarray}
or it can also be written as 
\begin{eqnarray}
    \mathcal{H}^{I=0} (x,\xi,t) &=& \mathcal{H}^q (x,\xi,t) + \mathcal{H}^{\bar{q}} (x,\xi,t), \\
   \mathcal{H}^{I=1} (x,\xi,t) &=& \mathcal{H}^q (x,\xi,t) - \mathcal{H}^{\bar{q}} (x,\xi,t),
\end{eqnarray}
where $\mathcal{H}^q (x,\xi,t)$ supports the regime $x \in \big[0,1\big]$, while $\mathcal{H}^{\bar{q}} (x,\xi,t) = -\mathcal{H}^q (-x,\xi,t)$ support the regime $x\in \big[-1 +\xi, \xi \big]$. The regime with $x \in \big[0,\xi\big]$ is the so-called Efremov-Radyushkin-Brodsky-Lepage (ERBL) regime, and $x \in \big[\xi,1 \big]$ is known as the Dokshitzer-Gribov-Lipatov-Altarelli-Parisi (DGLAP) regime.

In the forward limit, pion GPDs reduce to pion PDFs, corresponding to $\xi = 0$ and $t = 0$. In this kinematic limit, the initial and final pion momenta are equal, $p' = p$, and one obtains
\begin{eqnarray}
    \mathcal{H}^q (x,0,0) = \frac{1}{2} \big[ \mathcal{H}^{I=0} (x,0,0) + \mathcal{H}^{I=1} (x,0,0)\big] = u_v^\pi (x),
\end{eqnarray}
where $u_v^\pi(x) = u_\pi(x) - \bar{u}_\pi(x)$ denotes the valence quark distribution of the pion. In this expression, the valence quark distribution satisfies the normalization $\int_0^1 u_v^\pi (x)dx = 1$.

Based on the polynomiality condition, a general relation between the pion GPDs and the generalized form factors can be introduced by
\begin{eqnarray}
    \int_{-1}^{1} dx x^n \mathcal{H}^q (x, \xi,t) &=& \sum_{i=0}^{l} \xi^{2i} A_{n+1,2i}^q (t), \\
    \int_{-1}^{1} dx x^n  E^q (x,\xi,t)  &=& \sum_{i=0}^{l} \xi^{2i} B_{n+1,2i}^q (t), 
\end{eqnarray}
where $l= {(n+1)/2}$, and $A^q_{n+1} (t)$ and $B^q_{n+1} (t) $ are the coefficient functions of the form factors for the arbitrary values of $n$ and $i$. For $n=0$, it gives, respectively, the quark vector and tensor form factors. The vector and tensor form factors are determined from the twist-2 GPDs, while the scalar form factor is obtained from the twist-3 GPDs, and one has 
\begin{eqnarray}
\label{eq:piongffs}
    \int_{-1}^{1} dx \mathcal{H}^q (x,\xi=0,t) &=& A^q_{1,0} (t =-Q^2) = F^q_{\pi} (Q^2), \\
    \int_{-1}^{1} dx E^q (x,\xi=0,t) &=& B_{1,0}^q (t=-Q^2) = F^q_T (Q^2), \\
    \int_1^{-1} dx \mathcal{H}^q_{S} (x,\xi=0, t) &=& F^q_{S} (Q^2).
\end{eqnarray}

In the NJL model, the general expression of the vector, tensor, and scalar pion GPDs in Eq.~(\ref{eq:gpd1}) can be defined as 
\begin{eqnarray}
    \mathcal{H}^q (x,\xi,t) &=& 6ig_{\pi qq}^2 \int \frac{d^4k}{(2\pi)^4} \delta \Big( xP^+ - k^+ \Big)  \mathrm{Tr} \big[ \gamma_5 S_q (k) \gamma^+ S_q (k) \gamma_5 S_q (k-P) \big], \nonumber \\
    \mathcal{P} E^q (x,\xi,t) &=& 6i g_{\pi qq}^2  \int \frac{d^4k}{(2\pi)^4} \delta \Big( xP^+ -k^+ \Big) \mathrm{Tr} \big[ \gamma_5 S_q (k) i \sigma^{+j} S_q (k) \gamma_5 S_q (k-P)\big], \nonumber \\
  \frac{M_u}{P^+}  \mathcal{H}_{S}^q (x,\xi,t) &=& 6i g_{\pi q q}^2 \int \frac{d^4k}{(2\pi)^4} \delta \Big( xP^+ -k^+ \Big) \mathrm{Tr} \big[ \gamma_5 S_q (k) S_q (k) \gamma_5 S_q (k-P)\big], 
\end{eqnarray}
respectively, where $\mathcal{P}$ is defined in Eq.~(\ref{eq:gpd1}). After calculating the trace in the numerator, applying Feynman parameterization, Wick rotation, and introducing the proper time regularization scheme, the vector, tensor, and scalar for the up quark pion ($\pi^+ (u\bar{d})$) GPDs in the NJL model with $q=u$ are obtained as 
\begin{eqnarray}
    \label{eq:gpdnjl1}
    \mathcal{H}^u (x,\xi,t) &=& \frac{3 g_{\pi qq}^2}{8\pi^2} \Theta (\alpha_{0} ) \Theta ( \alpha_1 ) \int_{\tau_{\mathrm{UV}}^2}^{\tau_{\mathrm{IR}}^2} \frac{d\tau}{\tau} e^{\big[ -\tau \big( M_u^2 -\alpha_1 (1-\alpha_1) m_\pi^2\big) \big]} \nonumber \\
    &+& \frac{3g_{\pi qq}^2}{8\pi^2} \Theta (\alpha_2 )  \Theta ( \alpha_3)  \int_{\tau_{\mathrm{UV}}^2}^{\tau_{\mathrm{IR}}^2} \frac{d\tau}{\tau} e^{\big[ -\tau \big( M_u^2 - \alpha_2 (1-\alpha_2) m_\pi^2\big)  \big]} \nonumber \\
    &+& \frac{3xg_{\pi qq}^2 }{8\xi \pi^2} \Theta (\alpha_4)  \Theta (\alpha_5) \int_{\tau_{\mathrm{UV}}^2}^{\tau_{\mathrm{IR}}^2} \frac{d\tau}{\tau} e^{\big[ -\tau \big( M_u^2 - \beta_0 (1-\beta_0) t\big)  \big]} \nonumber \\
    &+& \frac{3g_{\pi qq}^2}{16\pi^2}  \Theta (\alpha_6) \Theta (\alpha_7) \big[(1-x)t + 2xm_\pi^2 \big] \nonumber \\
    &\times& \int_{\tau_{\mathrm{UV}}^2}^{\tau_{\mathrm{IR}}^2} d\tau \int_0^{1}d{\beta} \,\Theta\left(1-\beta-\beta_1\right) e^{\big[ -\tau \big( M_u^2 -{\beta} (1-{\beta})m_\pi^2 -\beta_1 (1-\beta_1 -{\beta} \big) t \big]}, ~~~
\end{eqnarray}
\begin{eqnarray}
    \label{eq:gpdnjl2}
    {E}^u (x,\xi,t) &=& \frac{3g_{\pi qq}^2 M_u m_\pi }{8\xi\pi^2}\Theta(\alpha_6)\Theta(\alpha_7) \nonumber\\
    &\times&\int_{\tau_{\mathrm{UV}}^2}^{\tau_{\mathrm{IR}}^2} d\tau\int_0^{1}d\beta \, \Theta\left(1-\beta-\beta_1\right) e^{\big[ -\tau \big( M_u^2 -{\beta} (1-{\beta})m_\pi^2 -\beta_1 (1-\beta_1 -{\beta} \big) t \big]}, ~~~
\end{eqnarray}
\begin{eqnarray}
\label{eq:gpdnjl3}
     \mathcal{H}^u_{S} (x,\xi,t) &=& \frac{3g_{\pi qq}^2}{4\pi^2}\Theta (\alpha_4)  \Theta (\alpha_5) \int_{\tau_{\mathrm{UV}}^2}^{\tau_{\mathrm{IR}}^2} \frac{d\tau}{\tau} e^{\big[ -\tau \big( M_u^2 - \beta_0 (1-\beta_0) t\big)  \big]}\nonumber\\
    &+&\frac{3g_{\pi qq}^2}{16\pi^2}\Theta (\alpha_6) \Theta (\alpha_7) \left[2m_\pi^2-t\right]\nonumber\\
    &\times&\int_{\tau_{\mathrm{UV}}^2}^{\tau_{\mathrm{IR}}^2} d\tau \int_0^{1}d{\beta} \, \Theta\left(1-\beta-\beta_1\right) e^{\big[ -\tau \big( M_u^2 -{\beta} (1-{\beta})m_\pi^2 -\beta_1 (1-\beta_1 -{\beta} \big) t \big]},~~~
\end{eqnarray}
where the variables in Eqs.~(\ref{eq:gpdnjl1})-(\ref{eq:gpdnjl3}) are defined as
\begin{align*}
    \alpha_0 &= \frac{x+\xi}{1+\xi}, & \alpha_1 &= \frac{1-x}{1+\xi}, \\
    \alpha_2 &= \frac{x-1}{\xi-1}, & \alpha_3 &= \frac{\xi-x}{\xi-1}, \\
    \alpha_4 &= 1-\frac{x}{\xi}, & \alpha_5 &= 1+\frac{x}{\xi}, \\
    \alpha_6 &= \frac{\xi+x-(1+\xi)\beta}{\xi}, & \alpha_7 &= \frac{\xi-x+(1-\xi)\beta}{\xi}, \\
    \beta_0 &= \frac{1}{2}\left(1+\frac{x}{\xi}\right) = \frac{1}{2}\alpha_5, & \beta_1 &= \frac{1}{2}\left(\frac{(1-\xi)\beta+\xi-x}{\xi}\right) = \frac{1}{2}\alpha_7.
\end{align*}
Hereafter, we use the expression for the $\mathcal{H}^u (x,\xi,t)$, $E^u (x,\xi,t)$, and $\mathcal{H}_S^u (x,\xi,t)$ for the vector, tensor, and scalar pion GPDs, respectively, as we focus on the pion GPDs in this work.

\section{Numerical Results and Discussion} \label{sec:num}
In this section, we present the numerical results for the pion GPDs, PDFs, and generalized form factors, including the scalar, vector, and tensor form factors, as well as their corresponding charge radii derived from the pion GPDs. In the computation of the pion GPDs, we use a dynamical quark mass of $M_u = M_{\bar{d}} = 400~\mathrm{MeV}$, assuming SU(2) isospin symmetry, together with an infrared cutoff $\Lambda_{\mathrm{IR}} = 240~\mathrm{MeV}$. The remaining parameters, $G_\pi = 19.03~\mathrm{GeV}^{-2}$ and $\Lambda_{\mathrm{UV}} = 645~\mathrm{MeV}$, are determined by fitting the pion mass $m_\pi = 140~\mathrm{MeV}$ and the pion decay constant $f_\pi = 93~\mathrm{MeV}$, following Refs.~\cite{Hutauruk:2016sug,Hutauruk:2018zfk}.

\subsection{Vector, tensor, and scalar pion GPDs}
\begin{figure}[ht]
\centering
\includegraphics[width=0.485\linewidth]{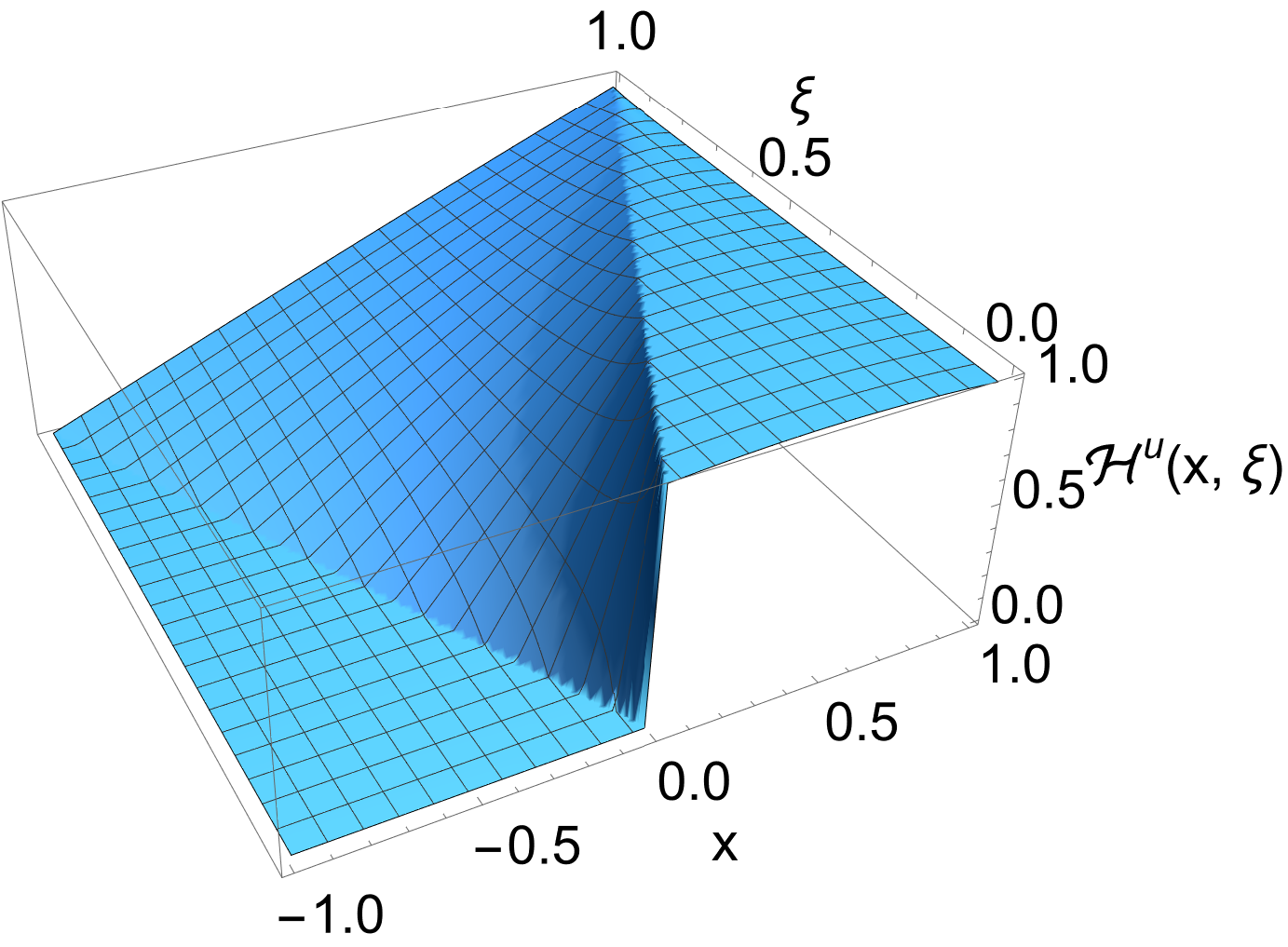}
\includegraphics[width=0.485\linewidth]{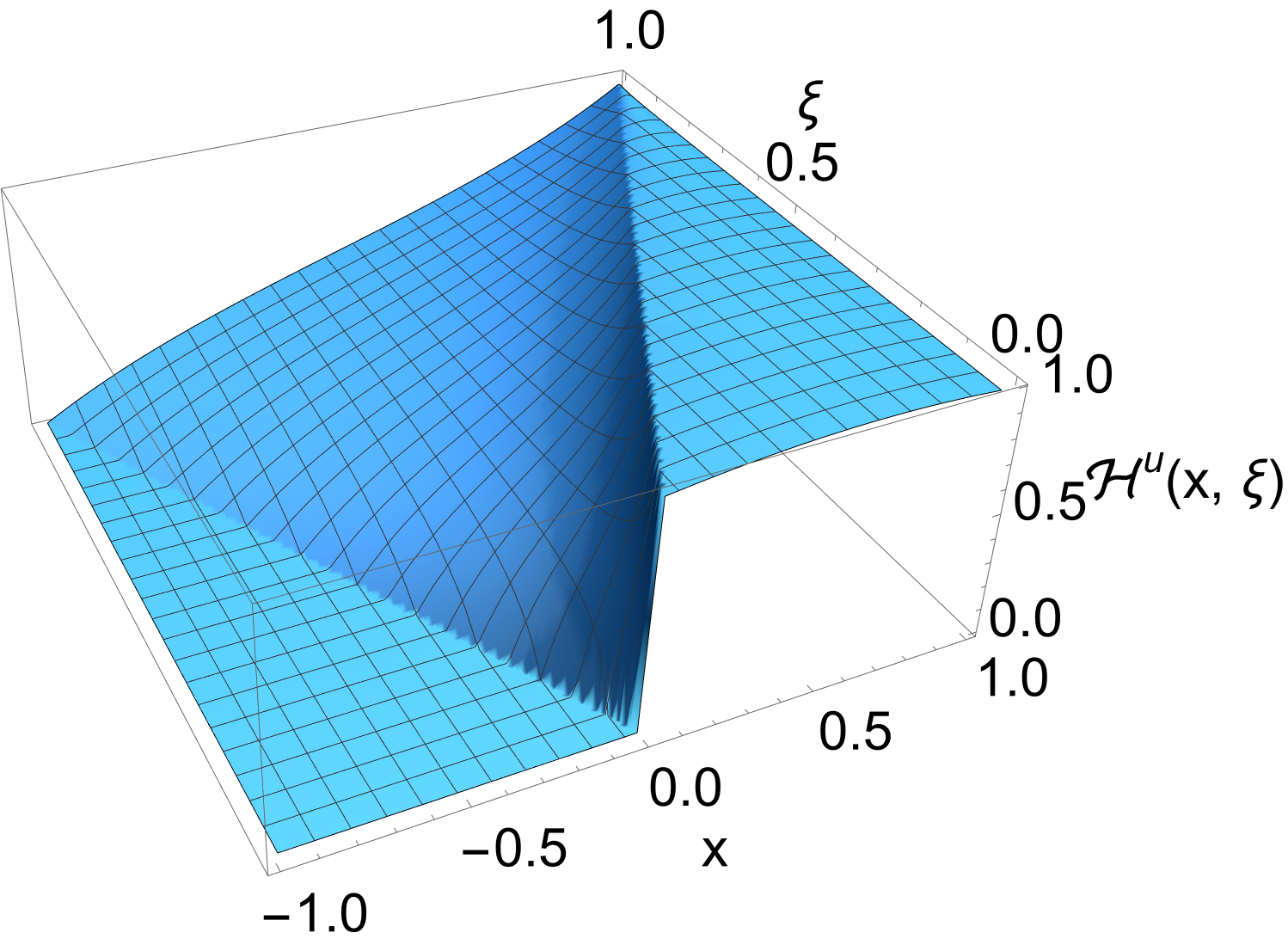} \\
\includegraphics[width=0.485\linewidth]{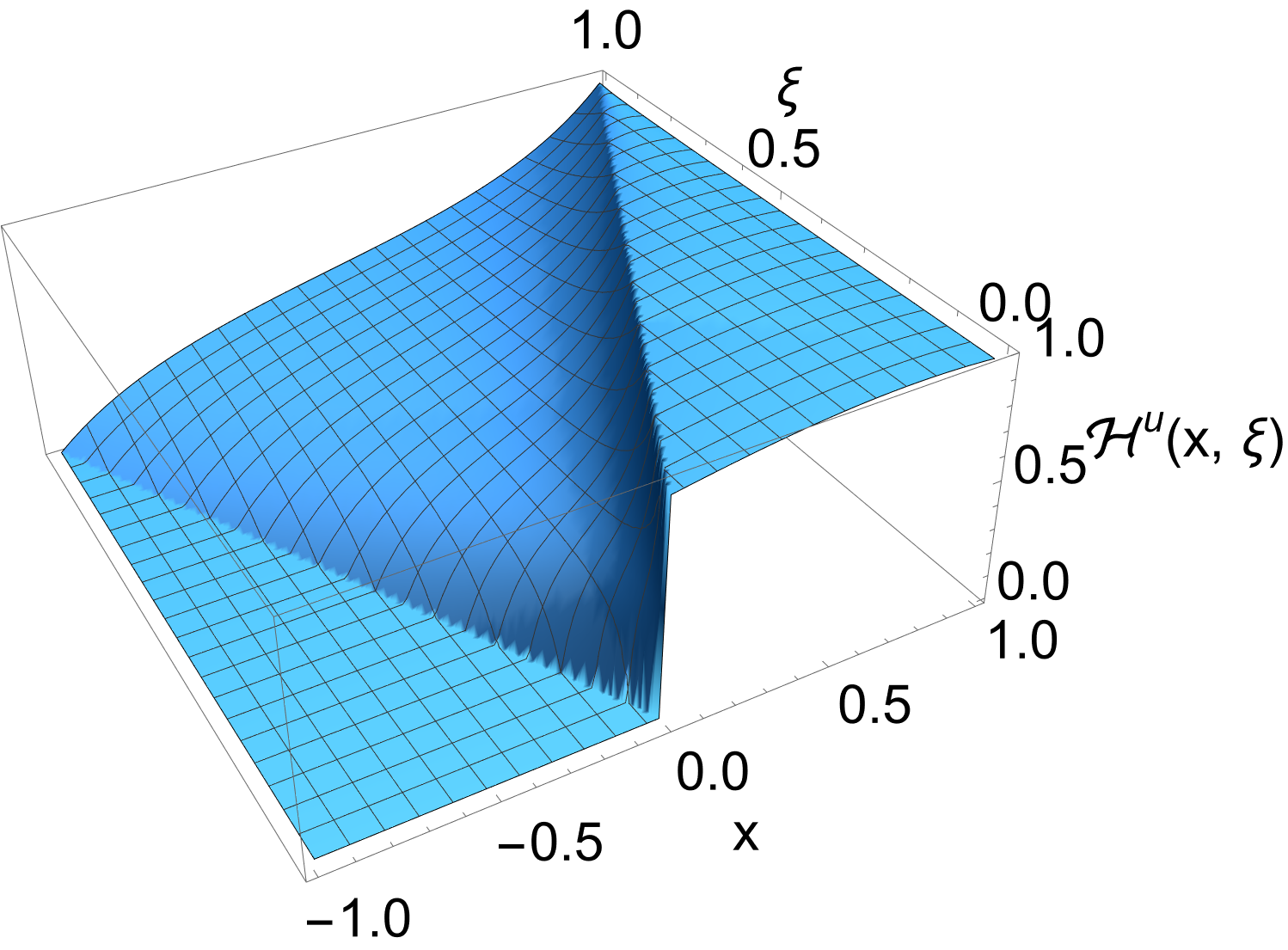}
\includegraphics[width=0.485\linewidth]{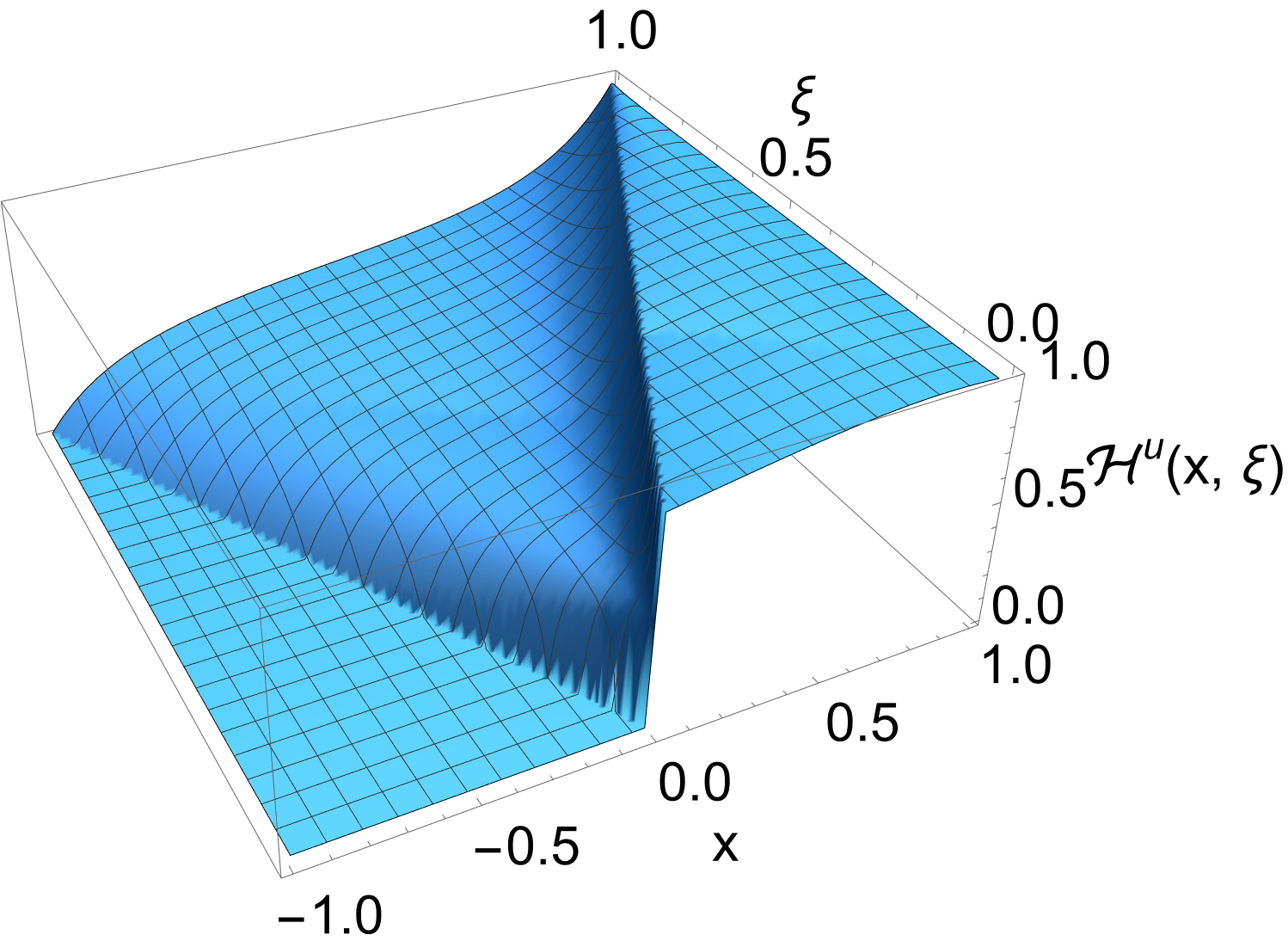}
\caption{The vector pion GPDs $\mathcal{H}^u (x,\xi,t)$ as functions of $\xi$ and $x$ for different $t =0.0$ GeV$^2$ (upper left panel), $-t=0.2$ GeV$^2$ (upper right panel), $-t=0.5$ GeV$^2$ (bottom left panel) and $-t=1.0)$ GeV$^2$ (bottom right panel).}
\label{gpd1}
\end{figure}

Our results for vector pion GPDs at $-t = 0.0$, $0.2$, $0.5$, and $1.0~\text{GeV}^2$ are shown in Fig.~\ref{gpd1}, illustrating how the shape of the vector pion GPDs evolves with increasing $t$. As expected, $\mathcal{H}^u(x,0,0)$ vanishes in the region $x \in [-1,0]$ for $\xi = 0$, while in the valence region $x \in [0,1]$ it equals unity. This can be attributed to the fact that the NJL model contains only valence quarks and no sea-quark (quark singlet) contributions at its intrinsic model scale. As momentum transfer $t$ increases, $\mathcal{H}^u(x,0,t)$ decreases, particularly in the vicinity of $x \simeq 0$.
\begin{figure}[ht]
\centering
\includegraphics[width=0.485\linewidth]{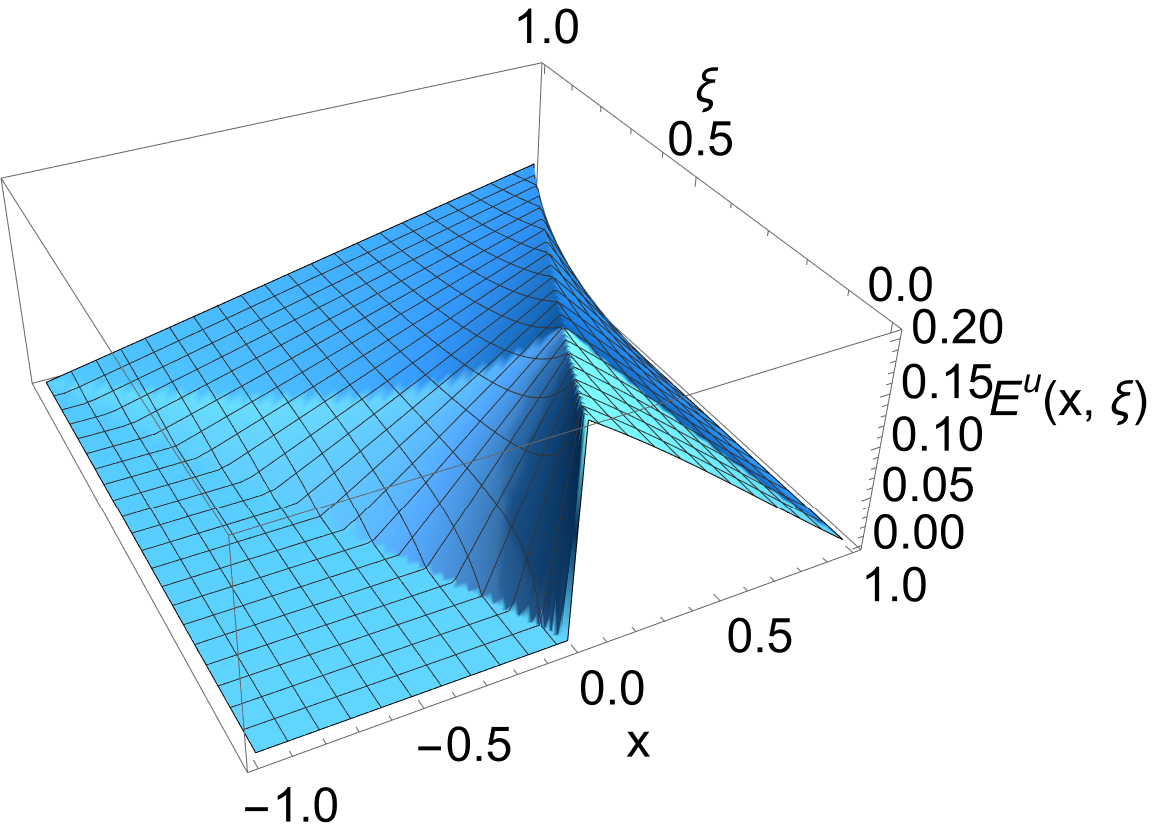}
\includegraphics[width=0.485\linewidth]{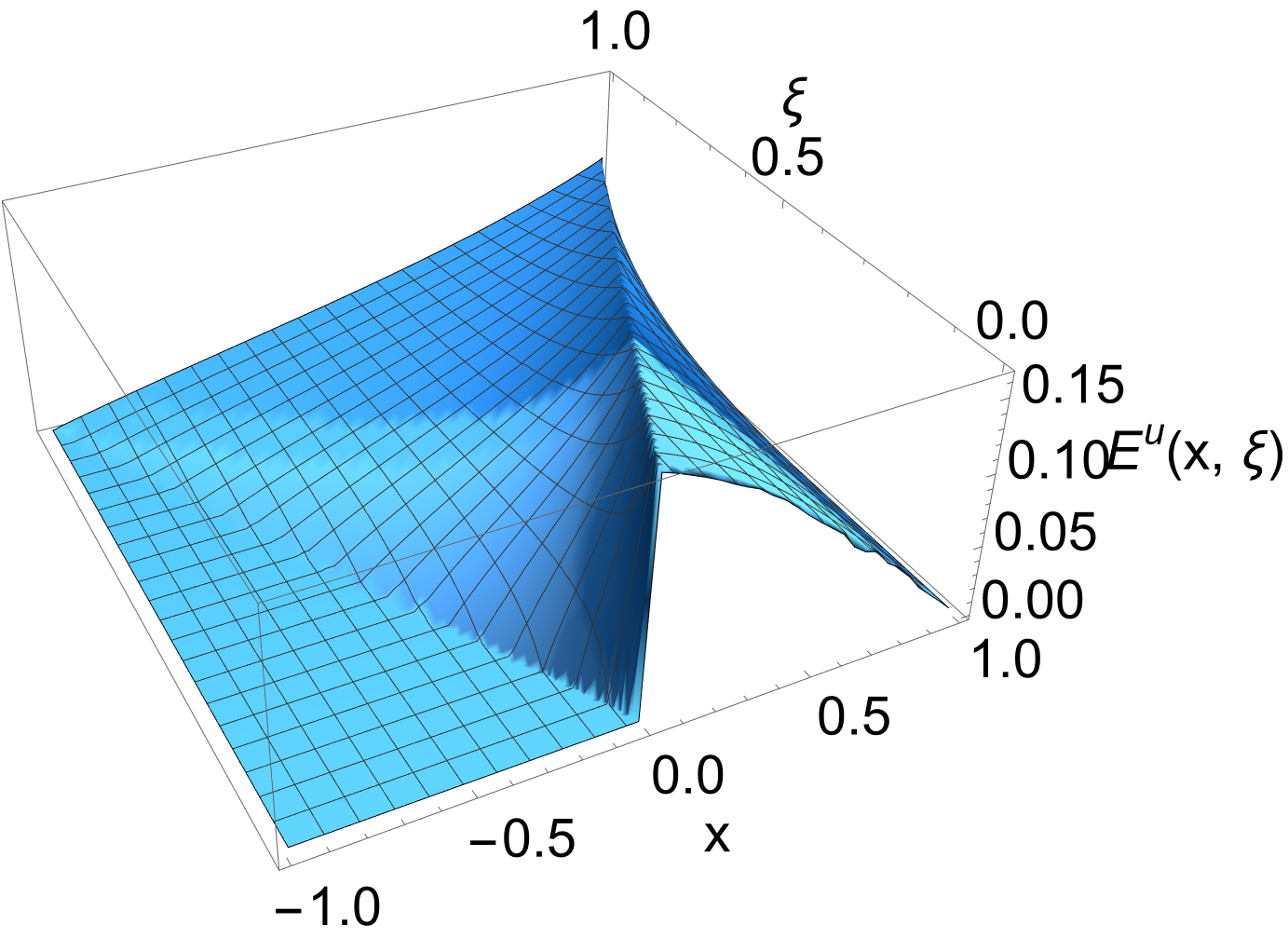} \\
\includegraphics[width=0.485\linewidth]{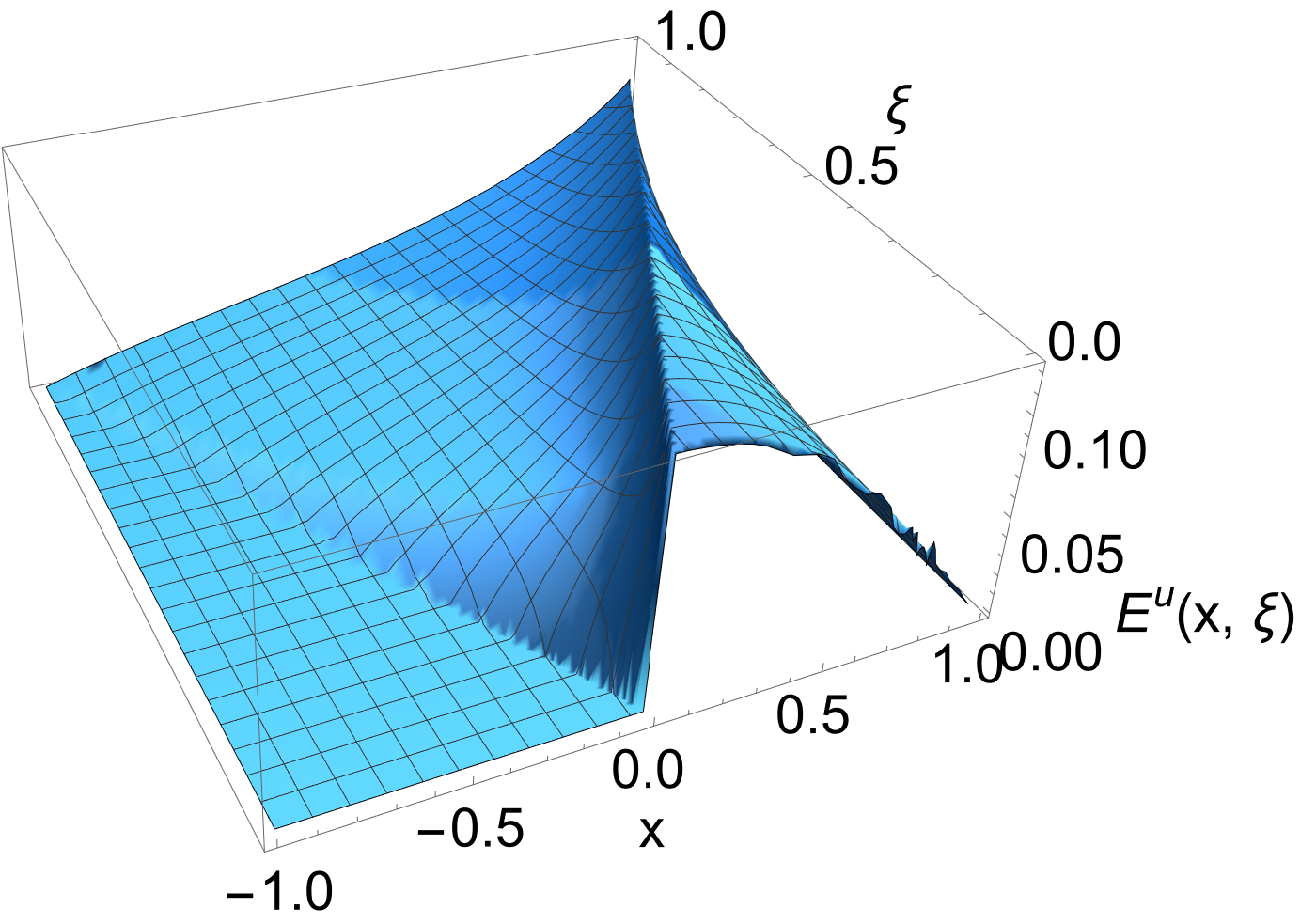}
\includegraphics[width=0.485\linewidth]{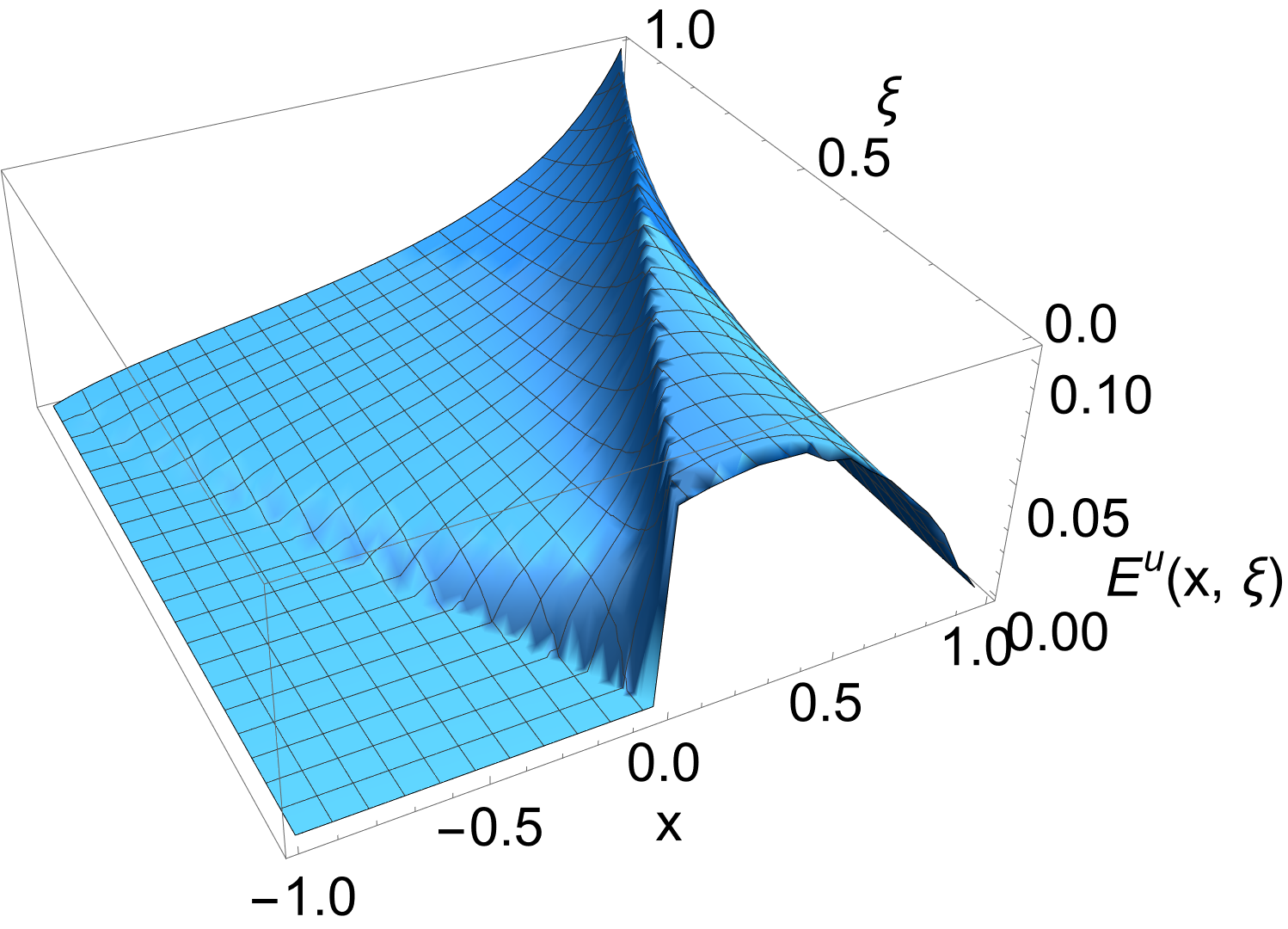}
\caption{Same as in Fig.~\ref{gpd1} but for the tensor pion GPDs $E^u (x,\xi,t)$.}
\label{gpd2}
\end{figure}

Analogous to the vector case, our results for the tensor pion GPDs at $-t = 0.0$, $0.2$, $0.5$, and $1.0~\text{GeV}^2$ are presented in Fig.~\ref{gpd2}. As in the vector pion GPDs, the shape of the tensor pion GPDs evolves with increasing $t$. For $\xi = 0$, we find that $E^u(x,0,0)$ vanishes in the region $x \in [-1,0]$. It is noteworthy that the peak of $E^u(x,0,0)$ at $t = 0$ occurs around $x \simeq 0.1$, and its position exhibits only a weak dependence on $t$ as the momentum transfer increases.
\begin{figure}[ht]
\centering
\includegraphics[width=0.485\linewidth]{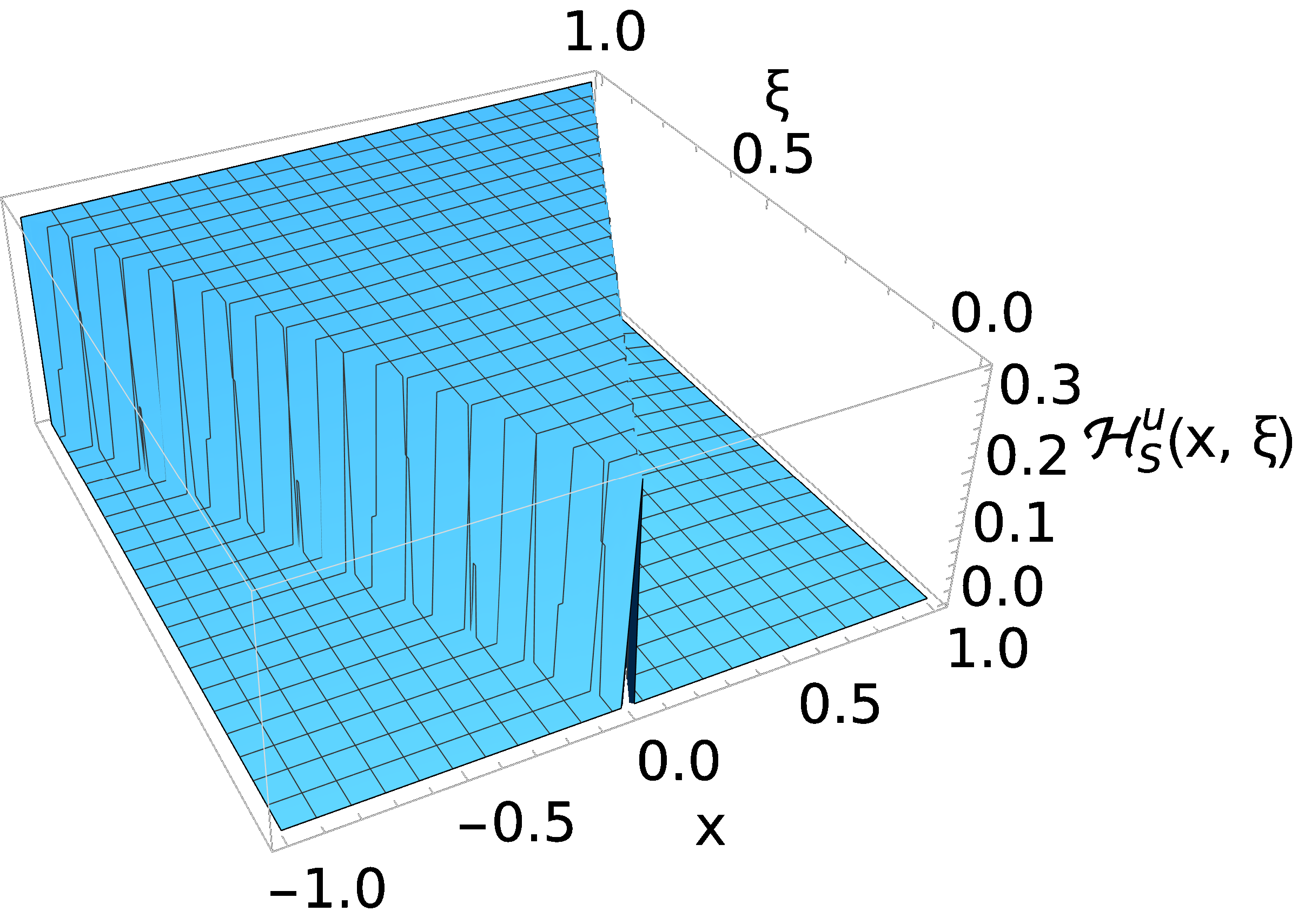}
\includegraphics[width=0.485\linewidth]{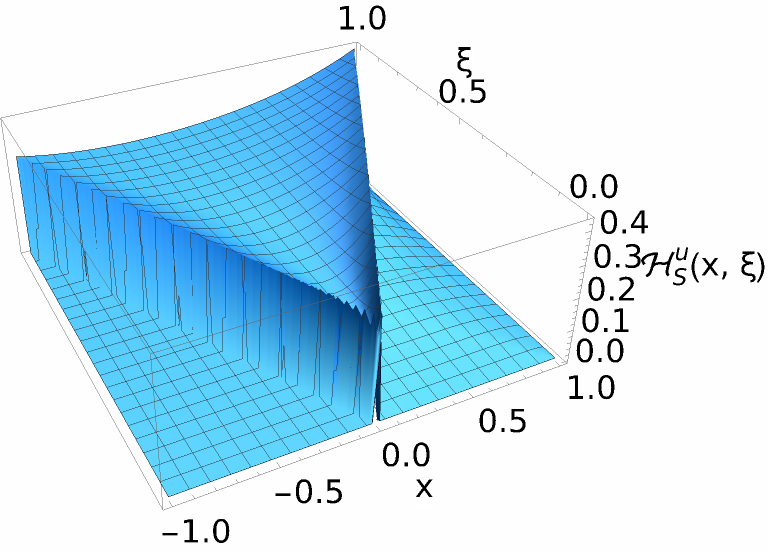} \\
\includegraphics[width=0.485\linewidth]{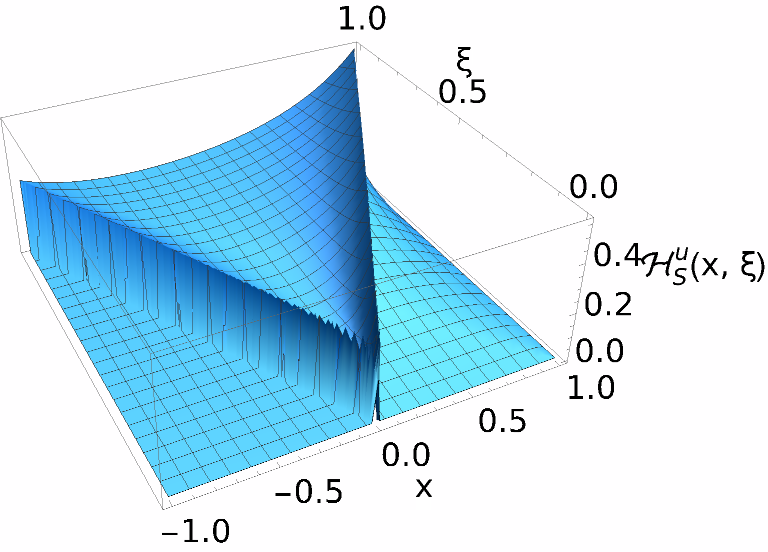}
\includegraphics[width=0.485\linewidth]{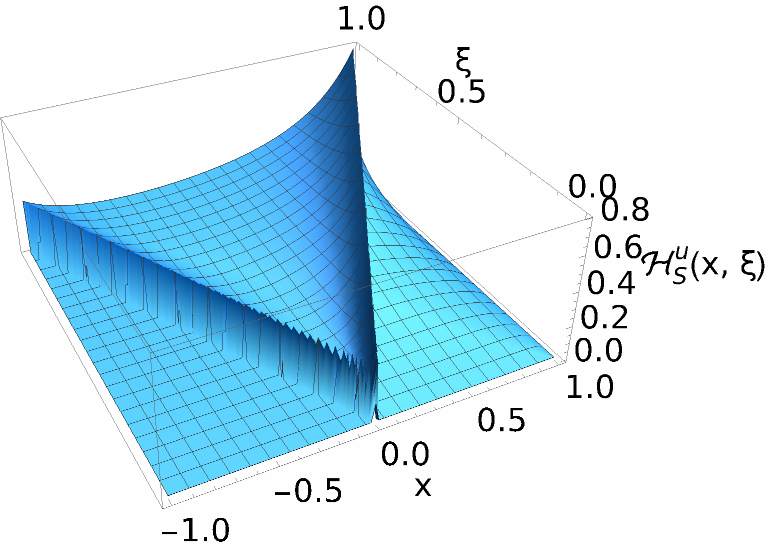}
\caption{Same as in Fig.~\ref{gpd1}, but for the scalar pion GPDs $\mathcal{H}_S^u (x,\xi,t)$.}
\label{gpd3}
\end{figure}

In addition to the results of the vector and tensor pion GPDs in Figs.~\ref{gpd1} and ~\ref{gpd2}, we also present our results for the twist-3 scalar pion GPDs for different values of $t$ in Fig.~\ref{gpd3}. The figure illustrates that the scalar pion GPDs also evolve with increasing momentum transfer. It is worth noting that the twist-3 scalar pion GPDs are computed in the same manner as the vector pion GPDs, except that the quark-photon vertex $\Gamma_{\gamma Q} = \gamma^+$ is replaced by $\Gamma_{\gamma Q} = 1$.

\subsection{Pion PDFs}
\begin{figure}[ht]
\centering
\includegraphics[width=0.485\linewidth]{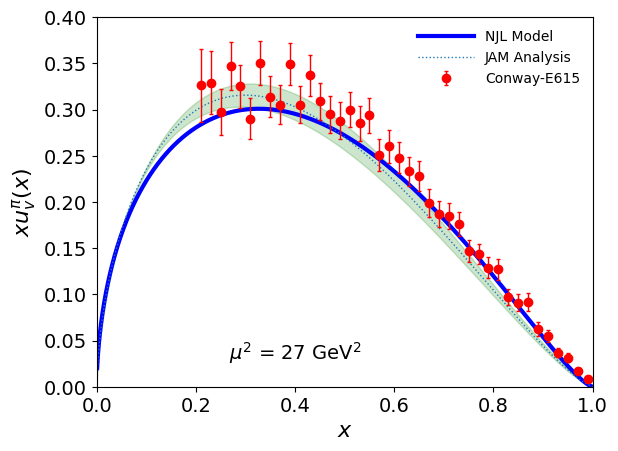}
\includegraphics[width=0.485\linewidth]{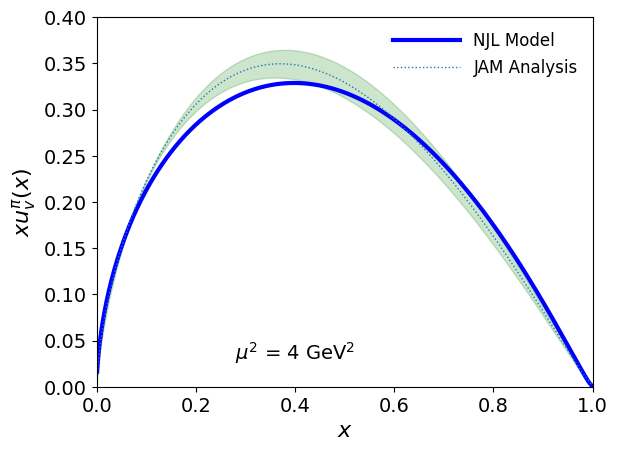}
\caption{Pion PDFs at $\mu^2 = $ 27 (left panel) and 4  (right panel) GeV$^2$. The pion PDFs are evolved from the initial model scale $\mu_0^2 = 0.18$ GeV$^2$. Note that the pion PDFs are computed from the pion GPDs. Experimental and JAM data are taken from Refs.~\cite{E615:1989bda}, and \cite{Barry:2021osv}, respectively.}
\label{pdf1}
\end{figure}

The pion PDFs [$\mathcal{H}^u (x,0,0) = u_v^\pi (x)$] in the NJL model can be computed from the vector pion GPDs $\mathcal{H}^u (x, \xi, t)$ by taking $\xi =0$ and $t=0$ (in the forward limit). Results for the pion PDFs at $\mu^2 =$ 27 GeV$^2$ and $\mu^2 =$ 4 GeV$^2$ in comparison with experimental data~\cite{E615:1989bda} and the JAM QCD analysis~\cite{Barry:2021osv} are shown in Fig.~\ref{pdf1}.

In the left panel of Fig.~\ref{pdf1}, we show the valence-quark pion PDFs ($x\mathcal{H}^u (x,0,0)$)  at $\mu^2 =$ 27 GeV$^2$, which is chosen based on the renormalization scale value of experimental data. We find that the valence-quark pion PDFs obtained from the pion GPDs are in excellent agreement with the existing experimental data~\cite{E615:1989bda} and the JAM analysis result~\cite{Barry:2021osv}. It is worth noting that this result validates our approach used in this work. With this result, we then predict the valence-quark pion PDFs at $\mu^2 =$ 4 GeV$^2$, where the values of the renormalization scale are adapted from the scale used in the lattice QCD and JAM analysis. 

The right panel of Fig.~\ref{pdf1} shows the results for the valence-quark pion PDFs in comparison with the JAM analysis result~\cite{Barry:2021osv} at renormalization scale $\mu^2 =$ 4 GeV$^2$. It is found that the valence-quark pion PDFs show excellent agreement with those obtained from the JAM analysis~\cite{Barry:2021osv}. Explicitly, it shows that the behavior of the quark power counting rule of the valence-quark pion PDFs at the asymptotic region ($ x\sim 1$) is consistent with the JAM analysis, expecting $xu_v^\pi (x) = x\mathcal{H}^u (x,0,0) \simeq (1-x)^{1}$ both at $\mu^2 = $ 27 and 4 GeV$^2$, respectively. This finding is slightly different from the DSE model prediction~\cite{Chang:2014gga} and the nonlocal chiral quark model (NLChQM)~\cite{Hutauruk:2025wkn,Hutauruk:2023ccw}, which predict a $(1-x)^{n}$ behavior with $n\simeq 2$ as $x \rightarrow 1$, as reported in Ref.~\cite{Hutauruk:2023ccw}. This difference is expected since the latter two models employ the momentum-dependent approach.


\subsection{Pion generalized form factors}
\begin{figure}[ht]
\centering
\includegraphics[width=0.485\linewidth]{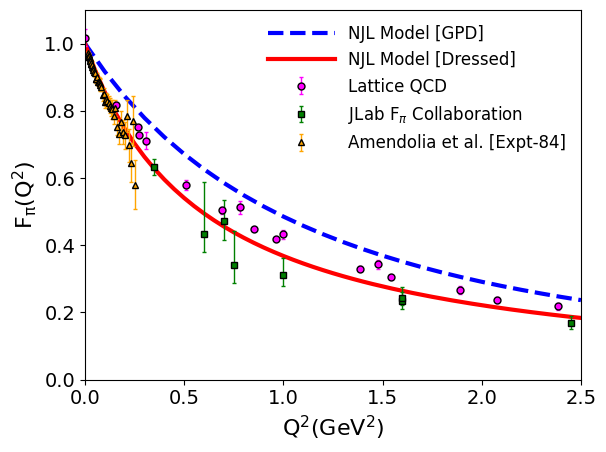}
\includegraphics[width=0.485\linewidth]{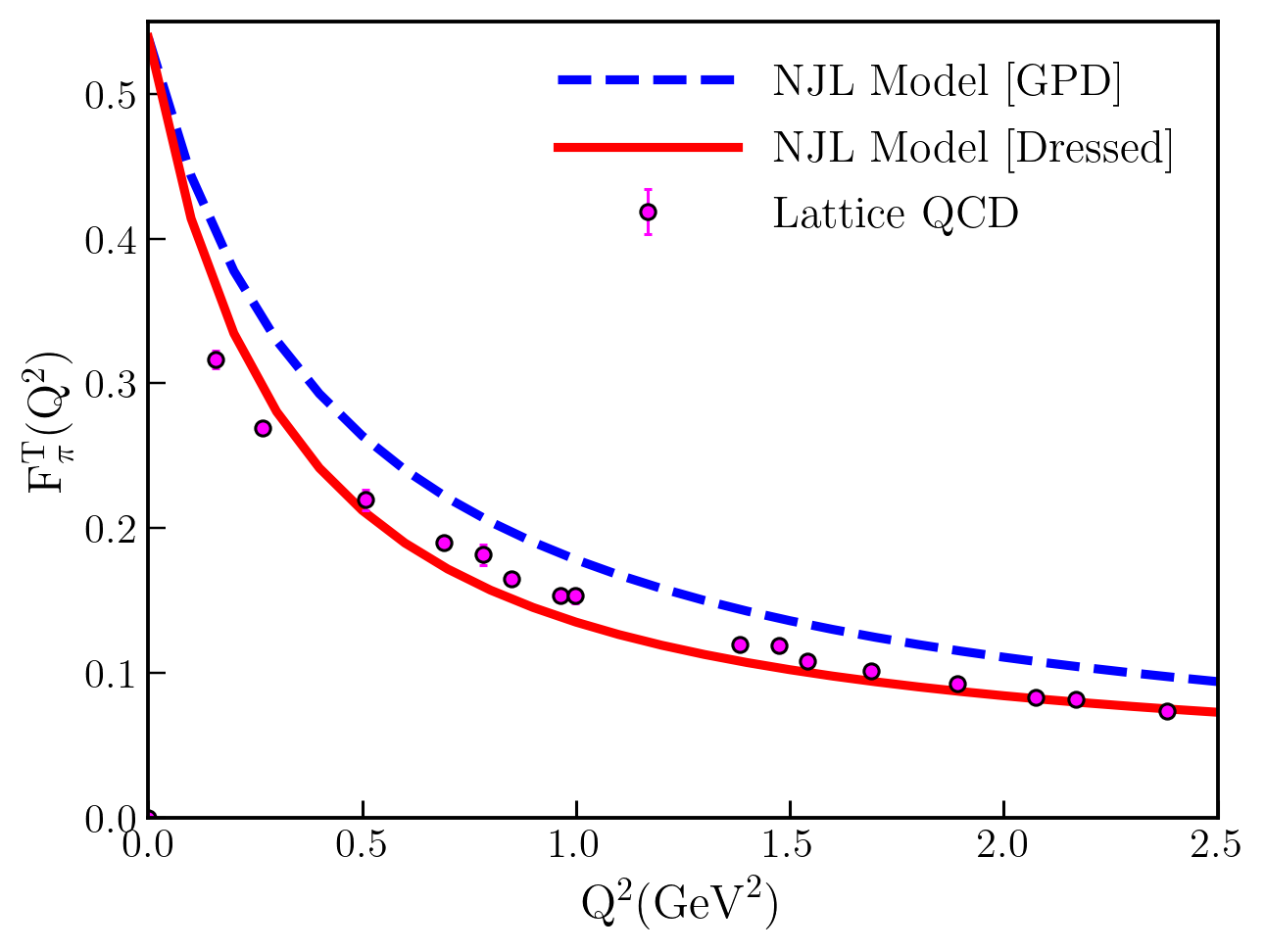}
\includegraphics[width=0.485\linewidth]{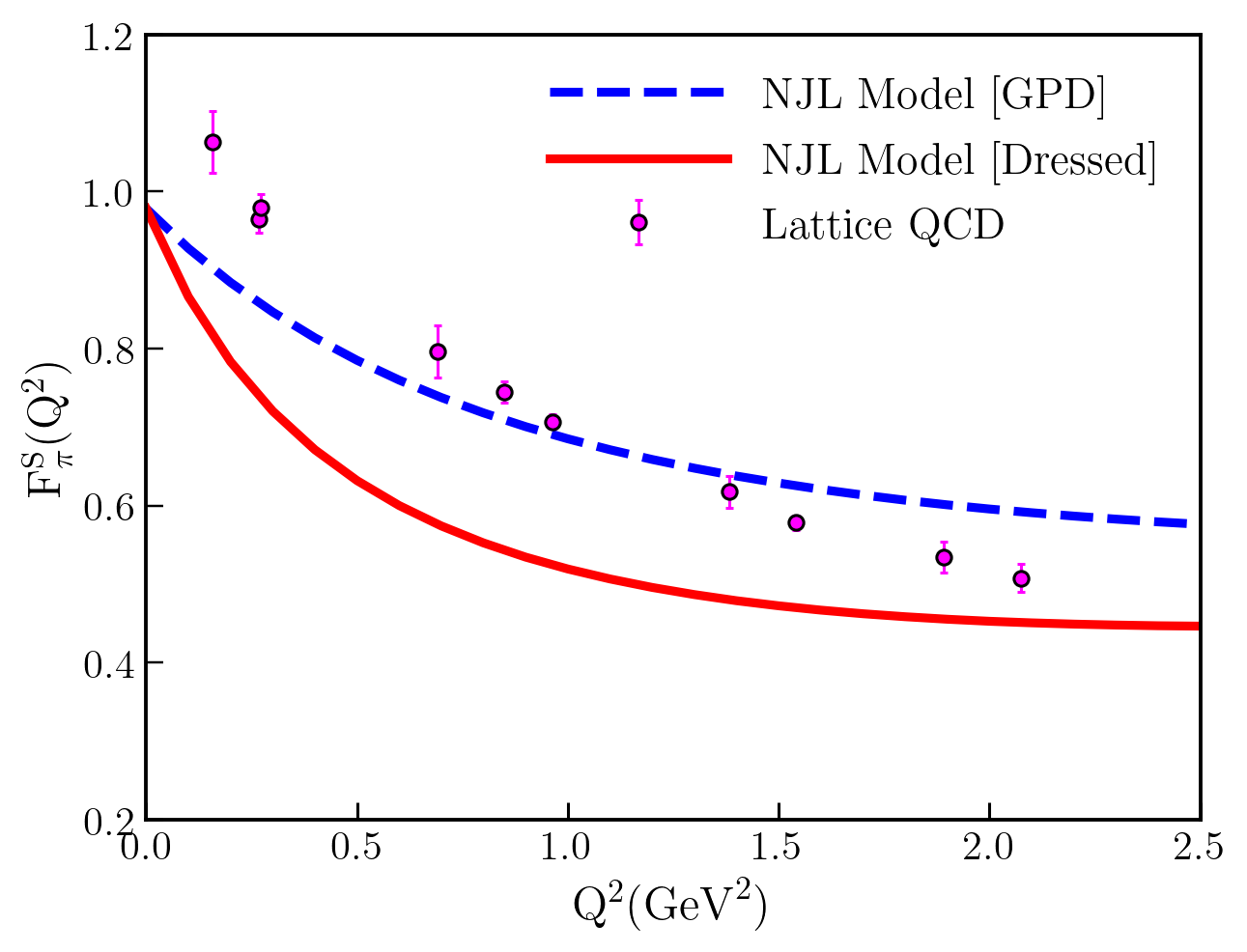}
\caption{The vector pion form factors $F^V_\pi (Q^2) = A_{1,0} (t)$ (upper left panel), scalar pion form factors $F^S_\pi (Q^2)$ (bottom center panel), and tensor pion form factors $F^T_\pi (Q^2) = B_{1,0} (t)$ (upper right panel) in comparison with the recent lattice QCD~\cite{Alexandrou:2021ztx} and experimental data~\cite{Amendolia:1984nz,NA7:1986vav,JeffersonLabFpi-2:2006ysh,JeffersonLabFpi:2007vir,JeffersonLab:2008jve,JeffersonLab:2008gyl}. Note that each panel is plotted with a different vertical scale. Note that the lattice results are calculated at a scale of 2 GeV, while our form factor results are obtained at the initial scale of the model.}
\label{gff1}
\end{figure}

Here, we present our results for the scalar, vector, and tensor pion generalized form factors derived from the twist-3 scalar pion GPDs and the $n=0$ Mellin moments of the pion GPDs. The results for the vector [$A_{1,0}^u (-t=Q^2) = F^V_{\pi} (Q^2)$], tensor [$B_{1,0}^u (-t=Q^2) = F^T_\pi (Q^2)$], and scalar $F^S_\pi (Q^2)$ pion form factors are depicted in Fig.~\ref{gff1}. 
\begin{figure}[ht]
\centering
\includegraphics[width=0.485\linewidth]{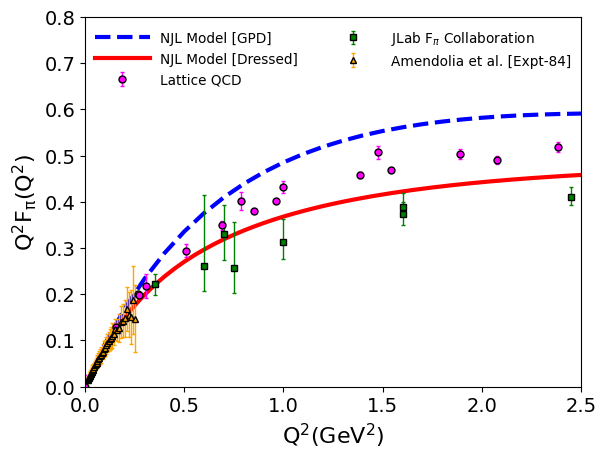}
\includegraphics[width=0.485\linewidth]{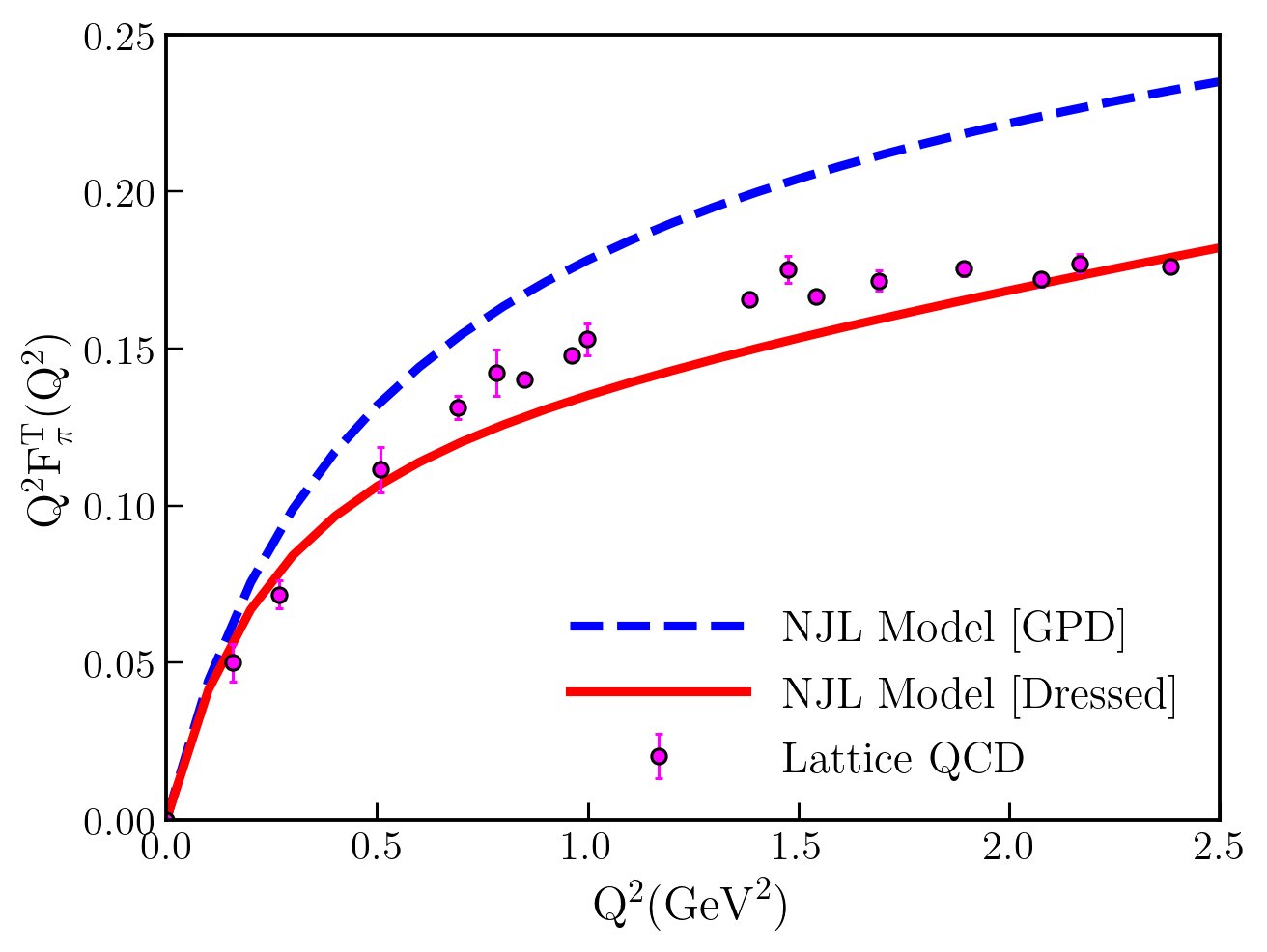}
\includegraphics[width=0.485\linewidth]{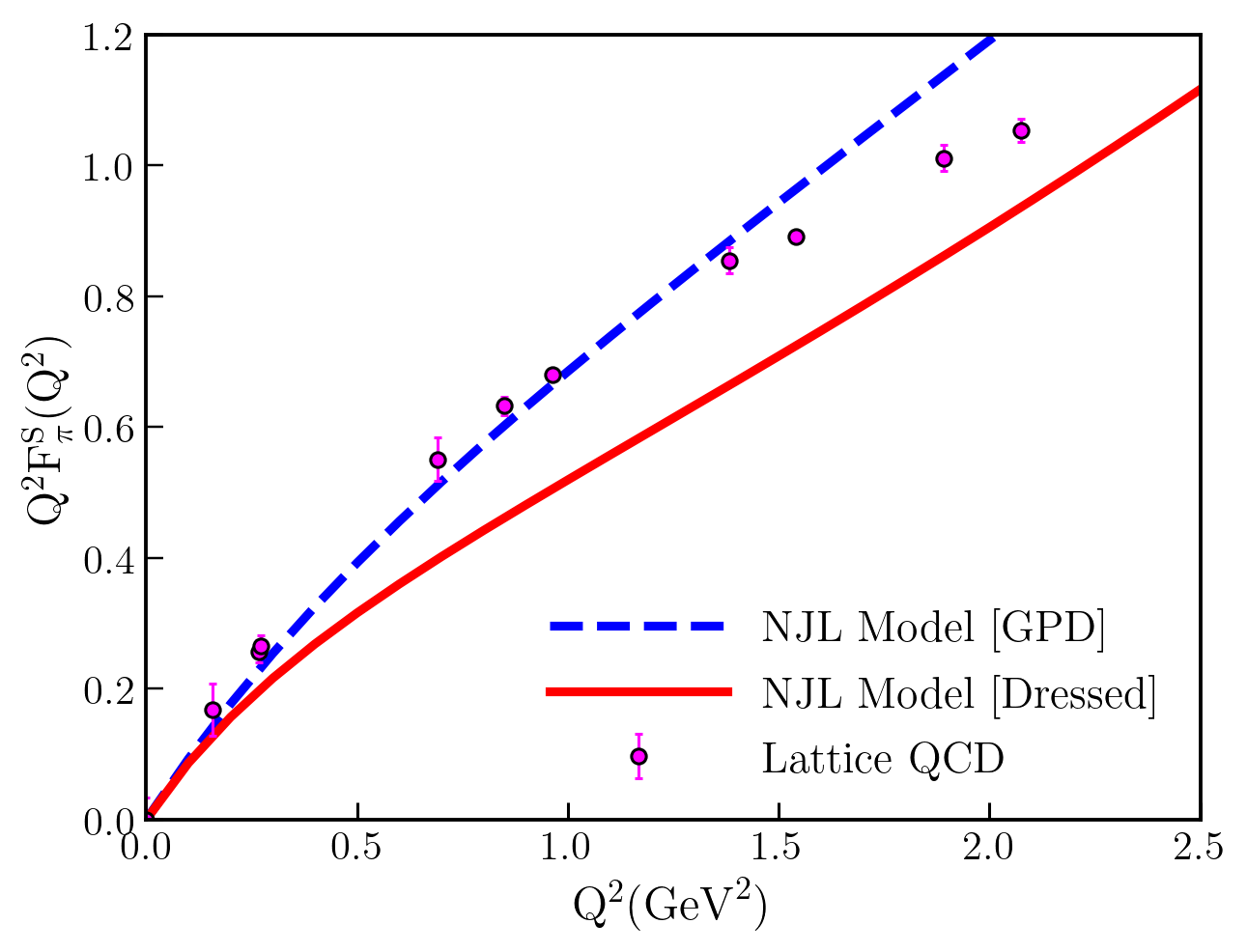}
\caption{Same as Fig.~\ref{gff1}, but for $Q^2F_\pi(Q^2)$ (upper left panel), $Q^2 F^S_\pi (Q^2)$ (bottom center panel) and $Q^2F_\pi^T(Q^2)$ (upper right panel). Note that each panel is plotted with a different vertical scale. Note that the lattice results are calculated at a scale of 2 GeV, while our form factor results are obtained at the initial scale of the model.}
\label{gff2}
\end{figure}

The upper left panel of Fig.~\ref{gff1} shows our results for the vector pion form factors $\left[F^V_\pi (Q^2) = F_\pi (Q^2)\right]$ obtained directly from the vector pion GPDs in comparison with the recent lattice QCD result~\cite{Alexandrou:2021ztx} and experimental data~\cite{Amendolia:1984nz,NA7:1986vav,JeffersonLabFpi-2:2006ysh,JeffersonLabFpi:2007vir,JeffersonLab:2008jve,JeffersonLab:2008gyl}. We find that our results for the bare vector pion form factor (blue dashed line) derived from the vector pion GPDs overestimate the recent lattice QCD~\cite{Alexandrou:2021ztx} and experimental data~\cite{Amendolia:1984nz,NA7:1986vav,JeffersonLabFpi-2:2006ysh,JeffersonLabFpi:2007vir,JeffersonLab:2008jve,JeffersonLab:2008gyl}. This is because the quark dressing in the quark–photon vertex has not yet been taken into account. The pion form factor obtained with the dressed quark is shown by the red solid line in Fig.~\ref{gff1}. The resulting dressed pion form factor is in excellent agreement with both the lattice data~\cite{Alexandrou:2021ztx} and the experimental measurements~\cite{Amendolia:1984nz,NA7:1986vav,JeffersonLabFpi-2:2006ysh,JeffersonLabFpi:2007vir,JeffersonLab:2008jve,JeffersonLab:2008gyl}.

In the upper right panel of Fig.~\ref{gff1}, we show our results for the tensor pion form factors in comparison with the recent lattice QCD result~\cite{Alexandrou:2021ztx}. We found that the bare tensor form factor of the pion overestimates recent lattice QCD results~\cite{Alexandrou:2021ztx}. Similar to the pion form factor, the dressed tensor pion form factor is in excellent agreement with that for the lattice data.

Result for the scalar pion form factors obtained from the twist-3 pion GPDs in comparison with the lattice QCD data is depicted in the bottom center of Fig.~\ref{gff1}. Surprisingly, the bare scalar pion form factor shows reasonable agreement with the lattice QCD data~\cite{Alexandrou:2021ztx} at intermediate values of $Q^2$, while the dressed scalar form factor underestimates the lattice QCD results~\cite{Alexandrou:2021ztx}.

In addition, we also present the corresponding results multiplied by the momentum transfer $Q^2$, as shown in Fig.~\ref{gff2}. The figure highlights the differences between our predictions and those from lattice QCD as well as experimental data. It also illustrates the behavior of the pion form factor (upper left panel), the tensor pion form factor (upper right panel), and the scalar pion form factor (bottom center panel) in the large-$Q^2$ (asymptotic) region. 

It is worth noting that the scalar, vector, and tensor pion form factors derived from the pion GPDs correspond to the bare form factors, which are not sufficient to describe experimental data and lattice QCD data. Consequently, dressing effects must be included in the form factors. In the dressed pion form factors, the quark-photon vertex is taken as $\Gamma_{\gamma Q} = \gamma^\mu F_{1\bar{Q}} (Q^2)$, where $\bar{Q} =(U, D)$ with $U$ and $D$ being the dressed of up and down quarks, respectively and $F_{1\bar{Q}} (Q^2)$ denotes the dressed quark form factors. A more comprehensive discussion in this regard is provided in Ref.~\cite{Hutauruk:2016sug}. 

\subsection{Charge radii}
Here, we present the charge radii associated with the vector, tensor, and scalar pion form factors. The corresponding scalar, vector, and tensor pion charge radii are extracted from the slopes of their respective form factors with respect to $Q^2$, and are given by
\begin{eqnarray}
      \big< r^{2}\big>_\pi^j &=& -\frac{6}{F_\pi^{j} (0)} \frac{\partial F_\pi^j (Q^2)}{\partial Q^2}\Bigg|_{Q^2=0}, 
\end{eqnarray}
where the superscripts $j=S, V$, and $T$ stand for the scalar, vector, and tensor pion charge radii, respectively.
\begin{figure}[ht]
    \centering
    \includegraphics[width=0.87\linewidth]{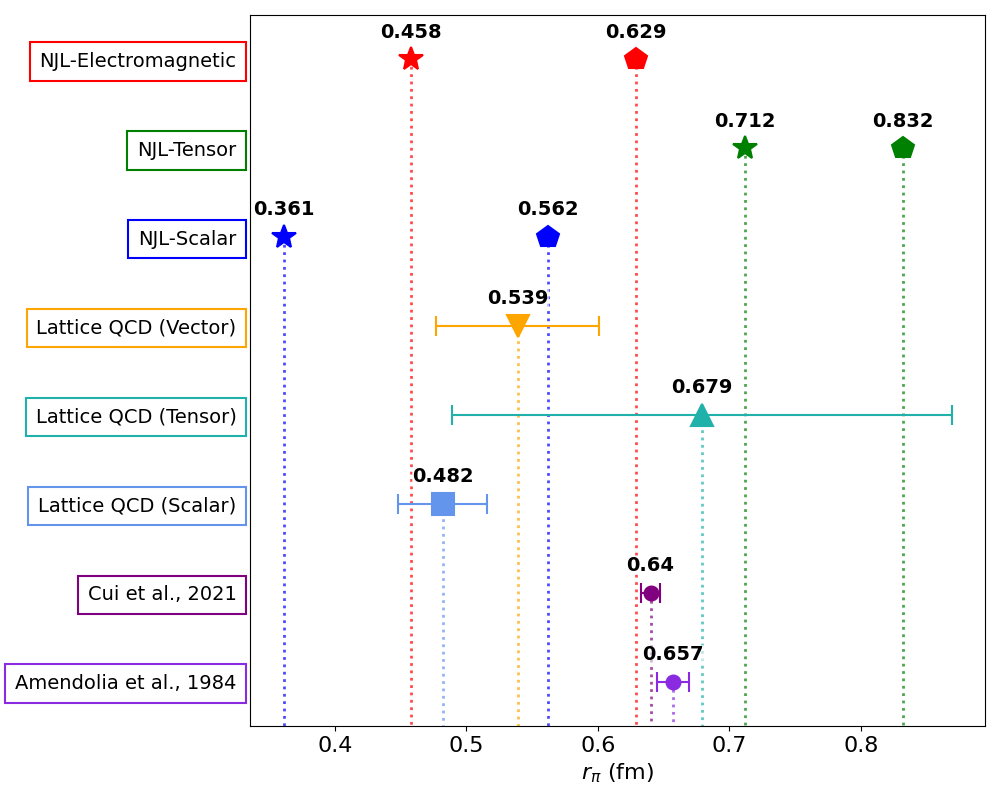}
    \caption{Charge radii for the vector, tensor, and scalar in comparison with experimental data, the lattice QCD data, and recent analysis for the pion form factor using the pion+electron elastic scattering data~\cite{Cui:2021aee}. Experimental data are taken from Ref.~\cite{Amendolia:1984nz}, while the lattice QCD (scalar), lattice QCD (vector), and lattice QCD (tensor) are obtained from Ref.~\cite{Alexandrou:2021ztx}. Note that the lattice results are calculated at a scale of 2 GeV, while our form factor results are obtained at the initial scale of the model.}
    \label{rad1}
\end{figure}

Results for the vector, tensor, and scalar pion charge radii in comparison with the lattice QCD calculations are shown in Fig.~\ref{rad1}. The values of the vector, tensor, and scalar charge radii for the pion, respectively, are obtained $r_V^\pi =$ 0.46 fm (bare), $r_T^\pi =$ 0.71 fm (bare), and $r_S^\pi =$ 0.36 fm (bare), while for the form factors with the dressed quark are $r_V^\pi =$ 0.63 fm (dressed), $r_T^\pi =$ 0.83 fm (dressed), and $r_S^\pi =$ 0.56 fm (dressed). We also find that the vector charge radius for the pion with dressed quark is consistent with experimental measurement~\cite{Amendolia:1984nz}, Particle Data Group (PDG) average~\cite{ParticleDataGroup:2016lqr}, lattice QCD result~\cite{Gao:2021xsm}, and other theoretical model calculations~\cite{Gifari:2024ssz,Hutauruk:2016sug}. A similar level of consistency is observed for the scalar pion charge radius when compared with lattice QCD results in Ref.~\cite{Alexandrou:2021ztx}, although our value is smaller than those obtained in the two-flavor lattice QCD and continuum results of Refs.~\cite{Gulpers:2013uca,Gulpers:2015bba}. For the tensor pion charge radius, we find that our result is reasonably consistent with the value reported in Ref.~\cite{Alexandrou:2021ztx}. Overall, we find that the order of the scalar, vector, and tensor pion charge radii is as follows: $r_T^\pi \geq r_V^\pi \geq r_S^\pi$, consistent with the order obtained in lattice QCD in Ref.~\cite{Alexandrou:2021ztx} and theoretical model calculations in Refs.~\cite{Puhan:2025pfs,Wang:2022mrh}.

\section{Summary and conclusion} \label{sec:sum}
To summarize, we have systematically investigated the parton distribution functions (PDFs) and generalized form factors (GFFs) of the pion, derived from the pion generalized parton distributions (GPDs) at zero skewness, within the covariant Nambu-Jona-Lasinio (NJL) model. The Schwinger proper-time regularization scheme was employed to control ultraviolet divergences and, at the same time, to simulate quark confinement. We computed the vector, tensor, and scalar pion GPDs and subsequently examined the pion PDFs obtained from the forward limit of the GPDs, as well as the vector and tensor pion form factors extracted from the first Mellin moments of the pion GPDs and the scalar pion form factor obtained from the twist-3 pion GPDs.

We found that the shapes of the scalar, vector, and tensor pion GPDs exhibit a mild dependence on the momentum transfer $t$, as shown in Figs.~\ref{gpd1}-\ref{gpd3}. Such variations are expected to influence the properties of the pion PDFs and GFFs in the forward limit. For the pion PDFs, we found that, after evolving from the initial scale $\mu_0^2 = 0.18~\text{GeV}^2$ to $\mu^2 = 27~\text{GeV}^2$, our results exhibit excellent agreement with the available experimental data~\cite{E615:1989bda} and the JAM global analysis~\cite{Barry:2021osv}. For completeness, we also evolved the pion PDFs to $\mu^2 = 4~\text{GeV}^2$ to enable a direct comparison with lattice QCD results and the JAM global analysis. At this scale, the resulting pion PDFs remain consistent with the JAM determination.

From the Mellin moment of the vector and tensor pion GPDs, we obtained the vector and tensor form factors, while the scalar pion form factor was derived from the twist-3 scalar pion GPDs. We found that the vector pion form factor is consistent with the experimental and lattice QCD data, followed by the tensor pion form factor, which is consistent with the lattice QCD data. Surprisingly, we found that the scalar pion form factor underestimates the lattice QCD data, as shown in the bottom center panel of Fig.~\ref{gff1}.

Finally, we computed the scalar, vector, and tensor pion charge radii. We obtained $r_V^\pi =$ 0.63 fm, $r_T^\pi =$ 0.83 fm, and $r_S^\pi =$ 0.56 fm. Using these values, we observed that the charge radii follow the ordering: $r_T^\pi \geq r_V^\pi \geq r_S^\pi$, which is consistent with the results reported by lattice QCD in Ref.~\cite{Alexandrou:2021ztx}, and theoretical model calculations in Refs.~\cite{Puhan:2025pfs,Wang:2022mrh}. It is worth emphasizing that direct experimental verification of the results presented in this study remains challenging. Nevertheless, these results may be accessible to future experimental programs, including the Electron-Ion Collider (EIC) at BNL~\cite{Arrington:2021biu}, the Electron-ion collider in China (EicC)~\cite{Anderle:2021wcy}, J-PARC~\cite{Sawada:2016mao}, the upgraded JLAB 22 GeV~\cite{Accardi:2023chb}, and the COMPASS/AMBER++ at CERN~\cite{Adams:2018pwt}.

\section*{Acknowledgments}
This work was supported by the PUTI Q1 Research Grant from the University of Indonesia (UI) under contract No. NKB 442/UN2.RST/HKP.05.00/2024 and the RCNP Collaboration Research Network program under project number COREnet 057.

\section*{ORCID}
\noindent Fernando Chandra \orcid{0009-0003-8583-4054}
\url{https://orcid.org/0009-0003-8583-4054}

\noindent Parada T.~P.~Hutauruk \orcid{0000-0002-4225-7109} \url{https://orcid.org/0000-0002-4225-7109}

\noindent Terry Mart \orcid{0000-0003-4628-2245} \url{https://orcid.org/0000-0003-4628-2245}


\end{document}